\documentclass[eqsecnum,showpacs,showkeys,a4]{revtex4}
\usepackage{amssymb}
\usepackage{graphicx}
\usepackage{bm}

\begin{document}

\title{Vector and axial-vector correlators in a nonlocal chiral quark model}
\author{Alexander E. Dorokhov}
\email{dorokhov@thsun1.jinr.ru}
\affiliation{Bogoliubov Laboratory of Theoretical Physics, Joint Institute for Nuclear
Research, 141980, Dubna, Russia,}
\affiliation{Instituto de Fisica Teorica, Universidade Estudial Paulista, 01405-900, Sao
Paulo, SP, Brazil}
\author{Wojciech Broniowski}
\email{b4bronio@cyf-kr.edu.pl}
\affiliation{The H. Niewodnicza\'nski Institute of Nuclear Physics, PL-31342 Cracow,
Poland}

\begin{abstract}
The behavior of nonperturbative parts of the isovector-vector and isovector
and isosinglet axial-vector correlators at Euclidean momenta is studied in
the framework of a covariant chiral quark model with nonlocal quark-quark
interactions. The gauge covariance is ensured with the help of the $P$%
-exponents, with the corresponding modification of the quark-current
interaction vertices taken into account. The low- and high-momentum behavior
of the correlators is compared with the chiral perturbation theory and with
the QCD operator product expansion, respectively. The $V-A$ combination of
the correlators obtained in the model reproduces quantitatively the ALEPH
data on hadronic $\tau $ decays, transformed into the Euclidean domain via
dispersion relations. The predictions for the electromagnetic $\pi ^{\pm
}-\pi ^{0}$ mass difference and for the pion electric polarizability are
also in agreement with the experimental values. The topological
susceptibility of the vacuum is evaluated as a function of the momentum, and
its first moment is predicted to be $\chi ^{\prime }(0)\approx (50\ \mathrm{%
MeV})^{2}$. In addition, the fulfillment of the Crewther theorem is
demonstrated.
\end{abstract}

\maketitle


\section{Introduction}

The vector ($V$) and axial-vector ($A$) current-current correlators are
fundamental quantities of the strong-interaction physics, sensitive to
small- and large-distance dynamics. They serve as an important testing
ground for QCD as well as for effective models of strong interactions. In
the limit of the exact isospin symmetry the $V$ and $A$ correlators in the
momentum space (with $-q^{2}\equiv Q^{2}\geq 0$) are defined as
\begin{eqnarray}
\Pi _{\mu \nu }^{V,ab}(q) &=&i\int d^{4}x~e^{iqx}\Pi _{\mu \nu
}^{V,ab}(x)=\,\left( q_{\mu }q_{\nu }-g_{\mu \nu }q^{2}\right) \Pi
_{T}^{V}(Q^{2})\delta ^{ab},  \label{PV} \\
\Pi _{\mu \nu }^{A,ab}(q) &=&i\int d^{4}x~e^{iqx}\Pi _{\mu \nu
}^{A,ab}(x)=\,\left( q_{\mu }q_{\nu }-g_{\mu \nu }q^{2}\right) \Pi
_{T}^{A}(Q^{2})\delta ^{ab}+q_{\mu }q_{\nu }\Pi _{L}^{A}(Q^{2})\delta ^{ab},
\label{PA} \\
\qquad \Pi _{\mu \nu }^{J,ab}(x) &=&\langle 0\left\vert T\left\{ J_{\mu
}^{a}(x)J_{\nu }^{b}(0)^{\dagger }\right\} \right\vert 0\rangle ,  \nonumber
\end{eqnarray}%
where the QCD currents are%
\begin{equation}
J_{\mu }^{a}=\overline{q}\gamma _{\mu }T^{a}q,\qquad J_{\mu }^{5a}=\overline{%
q}\gamma _{\mu }\gamma _{5}T^{a}q,  \label{JAV}
\end{equation}%
and $T^{a}$ denote the generators of the $SU_{F}(2)$ flavor group,
normalized to $\mathrm{tr}T^{a}T^{b}=\frac{1}{2}\delta ^{ab}$. The
momentum-space two-point correlation functions obey a (suitably subtracted)
dispersion relation,
\begin{equation}
\Pi (Q^{2})=\frac{1}{\pi }\int_{0}^{\infty }\frac{ds}{s+Q^{2}}\mathrm{Im}\Pi
(s),  \label{Peuclid}
\end{equation}%
where the imaginary parts of the correlators determine the spectral functions%
\begin{eqnarray}
v_{1}(s) &=&4\pi \mathrm{Im}\Pi _{T}^{V}(s+i0),  \nonumber \\
a_{1}(s) &=&4\pi \mathrm{Im}\Pi _{T}^{A}(s+i0).  \label{SpecDen}
\end{eqnarray}%
Recently, the inclusive nonstrange $V$ and $A$ spectral functions have been
determined separately and with high precision by the ALEPH \cite{ALEPH2} and
OPAL \cite{OPAL} collaborations from the hadronic $\tau $-lepton decays ($%
\tau \rightarrow \nu _{\tau }+$ hadrons) in the interval of invariant masses
up to the $\tau $ mass, $0\leq s\leq m_{\tau }^{2}$.

The difference of the $V$ and $A$ correlation functions is very sensitive to
the details of the spontaneous breaking of the chiral symmetry. In
particular, the behavior of this combination is constrained by the chiral
symmetry in the form of sum rules obtained through the use of the optical
theorem \cite{WIII,DMO,DGMLY,GasLeut85}. The experimental separation of the $%
V$ and $A$ spectral functions allows us to accurately test the chiral sum
rules in the measured interval \cite{ALEPH2,OPAL}. The coefficients of the
Taylor expansion of the correlators at $Q^{2}=0$ are expressed in terms of
the low-energy constants of the chiral perturbation theory($\chi $PT)\cite%
{GasLeut85}. On the other hand, the large-$s$ behavior of the correlators
can be confronted with perturbative QCD thanks to the sufficiently large
value of the $\tau $ mass. In the high-$s$ limit $\Pi ^{V}(Q^{2})$ and $\Pi
^{A}(Q^{2})$ are dominated by the free-field correlator, corrected by
nonperturbative terms with inverse powers of $Q^{2}$. This follows from the
fact that the correlators can be represented by an operator product
expansion (OPE) series and thus are sensitive to the nonperturbative physics
at smaller energy scales \cite{SVZ79}. Recently, the interest in the OPE
expansion has been revived due to a possible appearance of unconventional
quadratic power corrections, $\sim 1/Q^{2}$, found in \cite{CNZ}, and also
observed in lattice simulations \cite{Boucaud:2001st}. The ALEPH and OPAL
data have been intensely used in the literature in order to place limits on
the leading coefficients of the $\chi $PT and OPE expansions (see, \textit{%
e.g}, \cite%
{Davier:1998dz,ioffe,Geshkenbein:2002ri,SHURYAK,Klevansky:1997dk,deRafael:2002tj}%
).

The aim of this work is to calculate the nonperturbative parts of the $V$
and $A$ current-current correlators in the kinematic region reaching up to
moderately large Euclidean $Q^{2}$ and to extract experimentally observed
characteristics. The calculations are carried out in the effective chiral
model with nonlocal quark-quark interactions, which is made covariant by the
inclusion of the $P$-exponents in the non-local interaction vertex. A specific
prescription for the Wilson lines and their differentiation, described in
Sec. III, follows exactly Ref.~\cite{ADoLT00}. That way the model is made
consistent with the gauge invariance and can be used to analyze the $V$ and $%
A$ correlators. The model is a nonlocal extension of the well-known
Nambu-Jona Lasinio model. Moreover, its nonlocal structure may be motivated
by fundamental QCD interactions induced by the instanton and gluon
exchanges, which induce the spontaneous breaking of the chiral symmetry and
generate dynamically a momentum-dependent quark mass. From the point of view
of the standard OPE, the whole series of power corrections characterizes
nonlocal properties of the QCD vacuum and may be described in terms of the
nonlocal vacuum condensates \cite{MikhRad92,DEM97}. The use of a covariant
nonlocal low-energy quark model based on the self-consistent approach to the
dynamics of quarks has many attractive features, as it preserves the gauge
invariance, is consistent with the low-energy theorems, as well as takes
into account the large-distance dynamics controlled by the bound states.
Similar models with nonlocal four-quark interaction have been considered
earlier in, \emph{e.g.}, \cite%
{DP86,Tern91,IBGross92,Roberts:dr,Birse95,DoLT98,Bron99,bled,Golli,Prasza}
and applied to describe various low-energy phenomena.

Nonlocal models have an important feature which makes them advantageous over
the local models, such as the original Nambu--Jona-Lasinio model. At high
virtualities the quark propagator and the vertex functions of the quark
coupled to external fields reduce down to the free quark propagator and to
local, point-like couplings. This property allows us to straightforwardly
reproduce the leading terms of the operator product expansion. For instance,
the second Weinberg sum rule is reproduced in the model \cite{Bron99}, which
has not been the case of the local approaches. In addition, the intrinsic
nonlocalities, inherent to the model, generate unconventional power and
exponential corrections which have the same character as found in \cite{CNZ}
and in the instanton model (see \emph{e.g.} \cite{SHURYAK}). Recently, the
nonlocal effective model was successively applied to the description of the
data from the CLEO collaboration on the pion transition form factor in the
interval of the space-like momentum transfer squared up to 8 GeV$^{2}$ \cite%
{ADT99pigg,AD02}. Importantly, in that study at zero photon virtualities the
chiral anomaly result were reproduced, while at high photon virtualities the
factorization of short and long distances occurs at a scale of the order of
1~GeV$^{2}$. This allowed for the extraction of the pion distribution
amplitudes of leading and next-to-leading twists. There are several further
advantages in using the nonlocal models compared to the local approaches.
The non-local interactions regularize the theory in such a way that the
anomalies are preserved \cite{Ripka93,Arrio}. In other regularization
methods in the local models \cite{mitia:rev,Bijnens,Goeke96} the
preservation of the anomalies can only be achieved if the (finite) anomalous
part of the action is left unregularized, and only the non-anomalous
(infinite) part is regularized. Next, with non-local interactions the model
is finite to all orders in the $1/N_{c} $ (loop) expansion. Finally, as
shown in \cite{Golli}, stable solitons exist in a chiral quark model with
non-local interactions without the extra constraint that forces the chiral
fields to lie on the chiral circle.

In the present paper we further test the nonlocal quark model by carrying
out an analysis of the momentum dependence of the current-current
correlators. A transformation of the spectral functions measured by the
ALEPH collaboration into the Euclidean momentum space allows us for a
precise an unambiguous comparison of the experimental data with the model
calculations. The paper is organized as follows. In Sect. II, we briefly
recall the results of the chiral perturbation theory and operator product
expansion concerning the $V$ and $A$ correlators. In Sect. III and IV, we
outline the gauged nonlocal quark model and introduce the quark\--\-current
vertices. Then, we derive the expressions for the nonperturbative parts of
transverse $V$ and $A$ correlators (Sect. V) and, after fixing the model
parameters in Sect. VI, confront the results with the available experimental
data at large (Sect. VII) and low (Sect. VIII) Euclidean momenta. We
explicitly demonstrate the transverse character of the $V$ and nonsinglet $A$
correlators in Sect. IX. In Sect. X the contribution of the $U_{A}(1)$ axial
anomaly to the flavor-singlet longitudinal $A$ correlator is displayed and
the topological susceptibility is calculated as a function of the momentum.
Below, in all cases we use the strict chiral limit, with current quark mass
equal to zero. Numerically current quark mass corrections to observables
discussed in the paper do not exceed more than few percents.

\section{Chiral sum rules and the operator product expansion}

Chiral sum rules are dispersion relations between the real and absorptive
parts of a two-point correlation function that transforms symmetrically
under $SU(2)_{L}\times SU(2)_{R}$ (for the case of non-strange currents).
Through the use of the dispersion relations the sum rules are directly
expressed in terms of the difference of the $V$ and $A$ spectral densities.
Here is the list of sum rules, given in the strict chiral limit, which are
investigated in this paper:

\bigskip

\noindent The first Weinberg sum rule (WSR I) \cite{WIII},%
\begin{equation}
\frac{1}{4\pi ^{2}}\int_{0}^{s_{0}\rightarrow \infty }ds\left[ v_{1}\left(
s\right) -a_{1}\left( s\right) \right] =\left[ -Q^{2}\Pi _{T}^{V-A}\left(
Q^{2}\right) \right] _{Q^{2}\rightarrow 0}=f_{\pi }^{2},  \label{WsrI}
\end{equation}%
the second Weinberg sum rule (WSR II) \cite{WIII},
\begin{equation}
\frac{1}{4\pi ^{2}}\int_{0}^{s_{0}\rightarrow \infty }dss\left[ v_{1}\left(
s\right) -a_{1}\left( s\right) \right] =Q^{2}\left[ -Q^{2}\Pi
_{T}^{V-A}\left( Q^{2}\right) \right] _{Q^{2}\rightarrow \infty }=0,
\label{WsrII}
\end{equation}%
the Das-Mathur-Okubo (DMO) sum rule \cite{DMO},%
\begin{equation}
\frac{1}{4\pi ^{2}}\int_{0}^{s_{0}\rightarrow \infty }ds\frac{1}{s}\left[
v_{1}\left( s\right) -a_{1}\left( s\right) \right] =\left. \frac{\partial }{%
\partial Q^{2}}\left[ Q^{2}\Pi _{T}^{V-A}\left( Q^{2}\right) \right]
\right\vert _{Q^{2}\rightarrow 0}=f_{\pi }^{2}\frac{\left\langle r_{\pi
}^{2}\right\rangle }{3}-F_{A},  \label{DMO}
\end{equation}%
and, finally, the Das-Guralnik-Mathur-Low-Yuong (DGMLY) sum rule \cite{DGMLY}%
,%
\begin{equation}
-\frac{1}{4\pi ^{2}}\int_{0}^{s_{0}\rightarrow \infty }dss\ln \frac{s}{\mu
^{2}}\left[ v_{1}\left( s\right) -a_{1}\left( s\right) \right]
=\int_{0}^{\infty }dQ^{2}\left[ -Q^{2}\Pi _{T}^{V-A}\left( Q^{2}\right) %
\right] =\frac{4\pi f_{\pi }^{2}}{3\alpha }\left[ m_{\pi ^{\pm }}^{2}-m_{\pi
^{0}}^{2}\right] ,  \label{DGMLY}
\end{equation}%
where in the last equation $\alpha \approx 1/137$ is the fine structure
constant. In the chiral limit of massless quarks the DGMLY sum rule is independent of
the arbitrary normalization
scale, $\mu ^{2}$, thanks to WSR II. It was shown by Witten \cite{Witt83}
that the positive electromagnetic mass shift of the charged pions is a
consequence of the DGMLY sum rule combined with the positivity property of
the $V-A$ correlator,
\begin{equation}
-Q^{2}\Pi _{T}^{V-A}\left( Q^{2}\right) \geqslant 0,\qquad \mathrm{for}%
\qquad 0\leqslant Q^{2}\leqslant \infty .  \label{WittIneuq}
\end{equation}%
According to Witten, if the bare $u$ and $d$ quarks were massless and the
mass shift were negative, the charged pions would become tachyons
destabilizing the QCD vacuum.

Whereas WSR I and DMO are low-energy sum rules (in the sense that the
right-hand sides involve correlators at low momenta), and are reproduced in
most low-energy effective quark models, WSR II is a high-momentum sum rule.
In local models it is not reproduced, as discussed shortly. The DMGLY sum
rule collects contributions from the whole range of $Q^2$, both soft and
hard.

The left hand sides of the sum rules (\ref{WsrI})-(\ref{DGMLY}) have been
determined with the experimental data from \cite{ALEPH2} and \cite{OPAL},
with $s_{0}$ taken as the upper limit of the interval of the invariant mass
covered by the experiment. The right-hand sides of the sum rules are the
theoretical predictions as $s_{0}\rightarrow \infty $. The DMO sum rule
relates the derivative of $Q^{2}$ times the correlator to the square of the
pion decay constant $f_{\pi }=\left( 92.4\pm 0.3\right) \mathrm{MeV}$ \cite%
{PDG} obtained from the decays $\pi ^{-}\rightarrow \mu ^{-}\overline{\nu }%
_{\mu }$ and $\pi ^{-}\rightarrow \mu ^{-}\overline{\nu }_{\mu }\gamma $, to
the pion axial-vector form factor $F_{A}=0.0058\pm 0.0008$ for the radiative
decays $\pi ^{-}\rightarrow l^{-}\overline{\nu }_{l}\gamma $, and to the
pion charge radius-squared $\left\langle r_{\pi }^{2}\right\rangle =\left(
0.439\pm 0.008\right) $ \textrm{fm}$^{2}$ obtained from a one-parameter fit
to the space-like data \cite{NA7}.

The listed chiral sum rules provide important restrictions on the
correlators at low and high energies. The first Weinberg sum rule (\ref{WsrI}%
) fixes the normalization of correlators and holds in all variants of the
Nambu--Jona-Lasinio models, local or nonlocal. In general, the coefficients
of the Taylor expansion of the correlators at low Euclidean momenta are
given by the low energy constants of the strong chiral Lagrangian. The
second Weinberg sum rule (\ref{WsrII}) signals that the leading asymptotics
of the high $Q^{2}$ power expansion of the $V-A$ correlator essentially
starts from dimension $d=6$ term, and as such is valid in the nonlocal
versions of the effective chiral quark models \cite{Bron99}. In local models
WSR II involves on the right-hand side the large constituent quark mass
times quark condensate, $M_{q}<\bar{q}q>$, thus is violated badly.
In this regard the nonlocal models are highly rewarding.

More detailed short-distance, or large $Q^{2}$, properties of the
correlators are represented by the QCD operator product expansion \cite%
{SVZ79}. For the $V-A$ and $V+A$ combinations the OPE provides the following
leading-twist terms in the chiral limit:
\begin{eqnarray}
\Pi _{T}^{V-A}(Q^{2}) &=&\sum_{d=6,8...}\frac{O_{V-A}^{d}}{Q^{d}}=\frac{%
O_{V-A}^{6}}{Q^{6}}+\mathcal{O}(\frac{1}{Q^{8}}),  \label{PVmA} \\
\Pi _{T}^{V+A}(Q^{2}) &=&\sum_{d=0,2,4...}\frac{O_{V+A}^{d}}{Q^{d}}=
\label{PVpA} \\
&=&-{\frac{1}{4\pi ^{2}}}\left( 1+\frac{\alpha _{s}}{\pi }\right) \ln \frac{%
Q^{2}}{\mu ^{2}}-\frac{\alpha _{s}}{4\pi ^{3}}\frac{\lambda ^{2}}{Q^{2}}+%
\frac{1}{12}\frac{\left\langle \frac{\alpha _{s}}{\pi }\left( G_{\mu \nu
}^{a}\right) ^{2}\right\rangle }{Q^{4}}+\frac{O_{V+A}^{6}}{Q^{6}}+\mathcal{O}%
(\frac{1}{Q^{8}}),  \nonumber
\end{eqnarray}%
where the vacuum matrix elements of dimension $d=6$ operators are%
\begin{eqnarray}
O_{V-A}^{6} &=&\pi \alpha _{s}\left[ \left\langle \left( \overline{u}\gamma
_{\mu }\lambda ^{a}d\right) \left( \overline{d}\gamma _{\mu }\lambda
^{a}u\right) \right\rangle -\left\langle \left( \overline{u}\gamma _{\mu
}\gamma _{5}\lambda ^{a}d\right) \left( \overline{d}\gamma _{\mu }\gamma
_{5}\lambda ^{a}u\right) \right\rangle \right] , \label{OV-A}\\
O_{V+A}^{6} &=&-\pi \alpha _{s}\left[ \left\langle \left( \overline{u}\gamma
_{\mu }\gamma _{5}\lambda ^{a}d\right) \left( \overline{d}\gamma _{\mu
}\gamma _{5}\lambda ^{a}u\right) \right\rangle +\left\langle \left(
\overline{u}\gamma _{\mu }\lambda ^{a}d\right) \left( \overline{d}\gamma
_{\mu }\lambda ^{a}u\right) \right\rangle +\frac{2}{9}\sum_{i=u,d}%
\sum_{j=u,d,s,...}\left\langle \left( \overline{q_{i}}\gamma _{\mu }\lambda
^{a}q_{i}\right) \left( \overline{q_{j}}\gamma _{\mu }\lambda
^{a}q_{j}\right) \right\rangle \right] ,
\nonumber\end{eqnarray}%
with $\lambda ^{a}$ being the color $SU(3)$ matrices.

The ${V-A}$ correlator does not acquire any perturbative contribution in the
limit of massless quarks, hence it is sensitive entirely to the chiral
symmetry breaking parameters. Already at relatively small $Q^{2}$ the $d=6$
term dominates in the expansion of $\Pi _{V-A}$. On other hand, the sum of
the correlators, $\Pi _{V+A}$, supplied with small power corrections, is
close to the free-field result for distances up to 1~fm \cite{SHURYAK}. In
the expansion of $\Pi _{V+A}$ we also included $d=2$ term which violates the
original OPE expansion of \cite{SVZ79}. Motivation for inclusion this term into
consideration was given in \cite{CNZ}. In (\ref{PVmA}) and (\ref{PVpA}) we
do not explicitly show exponentially suppressed terms that may be induced by
existence of instantons \cite{SVZ79}. The magnitudes of the vacuum matrix
elements which appear in the OPE can not be fixed from the first principles
and are fitted to various hadronic observables. Clearly, their determination
is bound to carry experimental and theoretical uncertainties. Only a sign of
the $d=6$ term in the $V-A$ correlator is fixed by the Witten inequality: $%
O_{V-A}^{6}\leqslant 0$.

Different models are used to estimate the vacuum expectation values. The
standard approach in the calculation of the dimension $d=6$ matrix elements
suggested and used in original work \cite{SVZ79} was to explore the
factorization hypothesis, i.e. the saturation of the four-quark matrix
elements with the intermediate vacuum state. Under this assumptions for the
dimension $d=6$ matrix elements one gets%
\begin{equation}
\left[ O_{V-A}^{6}\right] ^{\mathrm{factor}}=-32\pi \alpha _{s}\left\langle \overline{%
q}q\right\rangle ^{2}/9,\qquad \left[ O_{V+A}^{6}\right] ^{\mathrm{factor}}=64\pi
\alpha _{s}\left\langle \overline{q}q\right\rangle ^{2}/81.  \label{O6factor}
\end{equation}%
However, some authors conclude that probably the factorization hypothesis is
violated by a factor of 2-3 \cite{Narison}. Moreover, quite different result
appears if one uses the instanton liquid model to calculate these matrix
elements \cite{Shuryak:1982I,Shuryak:82II}.

In this work, for comparison with other model results, we use the following
typical values of the condensates found via standard QCD sum rules without
and with the inclusion of the $d=2$ term:
\begin{equation}
\frac{\alpha _{s}}{\pi }\lambda ^{2}=0,\qquad \qquad \qquad \left\langle
\frac{\alpha _{s}}{\pi }\left( G_{\mu \nu }^{a}\right) ^{2}\right\rangle
=0.012\mathrm{~GeV}^{4},\qquad \alpha _{s}\left\langle \overline{q}%
q\right\rangle ^{2}=2.4\cdot 10^{-4}\mathrm{~GeV}^{6}\qquad \qquad \lbrack
7,11],  \label{OPEconSVZ}
\end{equation}%
\begin{equation}
\frac{\alpha _{s}}{\pi }\lambda ^{2}=-0.12\mathrm{~GeV}^{2},\qquad
\left\langle \frac{\alpha _{s}}{\pi }\left( G_{\mu \nu }^{a}\right)
^{2}\right\rangle =0.022\mathrm{~GeV}^{4},\qquad \alpha _{s}\left\langle
\overline{q}q\right\rangle ^{2}=5.8\cdot 10^{-4}\mathrm{~GeV}^{6}\qquad
\qquad \lbrack 42].  \label{OPEcondens}
\end{equation}%
The above values hold at a typical renormalization scale of about 1~GeV.

\section{Gauging nonlocal models}

In local theories, the gauge principle of the minimum action uniquely
determines the interaction of the matter fields with the gauge fields.
However, in nonlocal theories such an interaction may be introduced in
various ways, and its transverse part cannot be uniquely defined \cite%
{Mand62}. In order to obtain the nonlocal action in a gauge-invariant form
with respect to external fields $V$ and $A$, we define the delocalized quark
field, $Q$, with the help of the Schwinger gauge phase factor, a.k.a. the
Wilson line or the link operator,
\begin{equation}
Q(x,y)=P\exp \left\{ i\int_{x}^{y}dz_{\mu }\left[ V_{\mu }^{a}(z)+A_{\mu
}^{a}(z)\gamma _{5}\right] T^{a}\right\} q(y),\qquad \overline{Q}%
(x,y)=Q^{\dagger }(x,y)\gamma ^{0}.  \label{Qxy}
\end{equation}%
Here $V_{\mu }^{a}(z)$ and $A_{\mu }^{a}(z)$ are the external gauge vector
and axial-vector fields, respectively, and $P$ is the operator of ordering
along the integration path, with $y$ denoting the position of the quark and $%
x$ being an arbitrary reference point. The $P$ operator arranges the
matrices in each term of the expansion of the exponent from the left to the
right in the order determined by the point $z$ moving along the path from $x$
to $y$.

\begin{figure}[]
\includegraphics[height=2.5cm]{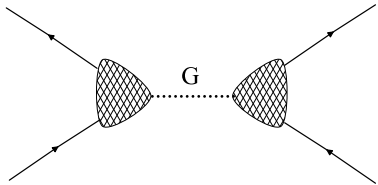}
\caption{{Diagrammatic representation of the effective nonlocal four-quark
interaction of Eq.~(\protect\ref{Lint}}). The hatched blobs represent the
nonlocal interactions, with the $\Gamma_i$ matrices present, and the dotted
line indicates that the diagram can be cut across this line without crossing
the quark lines going across the blobs.}
\label{w1}
\end{figure}

We start with the nonlocal chirally invariant action which describes the
interaction of soft quark fields. The nonlocal four-quark interaction is
depicted in Fig. \ref{w1}. The soft gluon fields have been integrated out.
The corresponding gauge-invariant action for quarks interacting through
nonperturbative exchanges can be expressed in a form similar to the
Nambu--Jona-Lasinio model \cite{ADoLT00}
\begin{eqnarray}
S &=&\int d^{4}x\ \overline{q}(x)\gamma ^{\mu }\left[ i\partial _{\mu
}-V_{\mu }\left( x\right) -\gamma _{5}A_{\mu }\left( x\right) \right] q(x)+
\nonumber \\
&+&\frac{1}{2}G\int d^{4}X\int \prod_{n=1}^{4}d^{4}x_{n}\ f(x_{n})\left[
\overline{Q}(X-x_{1},X)\Gamma _{i}Q(X,X+x_{3})\overline{Q}(X-x_{2},X)\Gamma
_{i}Q(X,X+x_{4})\right] ,  \label{Lint}
\end{eqnarray}%
where in the simplest version of the model the spin-flavor structure of the
interaction is given by matrix product
\begin{equation}
\left( \Gamma _{i}\otimes \Gamma _{i}\right) =\left( 1\otimes 1+i\gamma
_{5}\tau ^{a}\otimes i\gamma _{5}\tau ^{a}\right) .  \label{NJLnl}
\end{equation}%
In Eq.~(\ref{Lint}) $\overline{q}=(\overline{u},\overline{d})$ denotes the
quark flavor doublet field, $G$ is the four-quark coupling constant, and $%
\tau ^{a}$ are the Pauli isospin matrices.

\begin{figure}[]
\includegraphics[height=9.7cm]{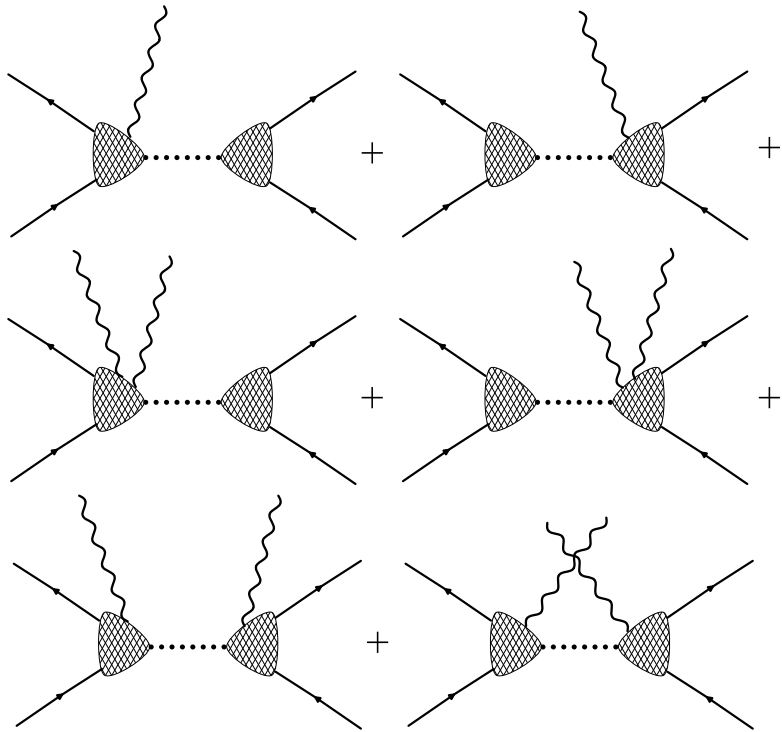}
\caption{{Diagrammatic representation of gauging of the effective nonlocal
four-quark interaction (\protect\ref{Lint}}), shown up to the second order
in the expansion in external fields, represented by the wavy lines. Note
that vertices involving one quark line and multiple gauge fields appear.}
\label{w2}
\end{figure}

The delocalization of the quark fields with the inclusion of the
path\--\-ordered Schwinger phase factors, Eq.~(\ref{Qxy}), ensures the gauge
invariance of the nonlocal action (\ref{Lint}). However, the presence of
these factors modifies the quark-current interaction, as shown graphically
in Fig. (\ref{w2}). The modification of interaction, required by the gauge
principle, poses a technical difficulty in dealing with nonlocal models, as
many diagrams appear in the analysis of physical processes. The ambiguities
in making the nonlocal 4-quark interaction gauge invariant are manifest in
the path-dependence in the definition (\ref{Qxy}), as well as in the choice
of the junction of the quark sources with the gauge strings. In general, the
Noether currents consist of two components: the path-independent
longitudinal part and the path-dependent transverse part. The dependence of
the transverse component on the choice of the path is a feature of any
method employed in constructing the Noether currents corresponding to a
nonlocal action, and this freedom is immanent to the formulation of the
model. We should recall here that the discussed ambiguities in the
construction of the transverse parts of the Noether currents are by no means
specific to chiral quark models. They also appear, \emph{e.g.}, in nuclear
physics when one considers meson-exchange processes. To summarize, the
choice of the path in Eq. (\ref{Qxy}) is a part of the model building.

In what follows, we use the formalism \cite{Mand62,Tern91} based on the
path-independent definition of the derivative of the integral over a line
for an arbitrary function $F_{\mu }(z)$:
\begin{equation}
\frac{\partial }{\partial y^{\mu }}\int_{x}^{y}dz^{\nu }\ F_{\nu }(z)=F_{\mu
}(y),\qquad \delta ^{(4)}\left( x-y\right) \int_{x}^{y}dz^{\nu }\ F_{\nu
}(z)=0.  \label{MinInt}
\end{equation}%
This choice effectively means that the differentiation involves moving the
end-point of the line only, with the other part of the line kept fixed. As a
result, the terms with nonminimal coupling, which are induced by the kinetic
term of the action, are omitted.

In general, external fields entering into Eq.~(\ref{Lint}) are arbitrary
auxiliary fields; however, some of them can be associated with
electromagnetic, weak, or strong interactions. In the case of the
electromagnetic interactions, the gauge factor takes into account the
effects of the radiation of the photon field when the two quarks are moving
apart. This formalism was used in \cite%
{Tern91,IBGross92,Birse95,Bron99,DoLT98} to represent the nonlocal
interaction in a gauge-invariant form. The incorporation of a
gauge-invariant interaction with gauge fields is of principal importance if
one desires to treat correctly the hadron characteristics probed by external
currents, such as hadron electromagnetic and weak form factors, structure
functions, distribution amplitudes, \emph{etc.}

In Eq.~(\ref{Lint}) the functions $f(x_{n})$, normalized to $f(0)=1$, form
the kernel of the four\--\-quark interaction and characterize the
space-dependence of the order parameter of the spontaneous chiral-symmetry
breaking. Thus, the interaction is treated in the separable approximation.
The choice of the nonlocal kernel in the form of (\ref{Lint}) is motivated
by the instanton-induced nonlocal quark-quark interaction \cite{DP86}, where
the nonlocal function $ f(x_{n})$ is related to the quark zero mode
emerging in the instanton field \cite{DP86,DEM97}. To have the same flavor
symmetry as in the original instanton-induced 't Hooft determinant
interaction one needs to add yet another piece of the form
\begin{equation}
G^{\prime }\left( \tau ^{a}\otimes \tau ^{a}+i\gamma _{5}\otimes i\gamma
_{5}\right) ,  \label{G5}
\end{equation}%
with the coupling $G^{\prime }=-G$. This term will be important in the
discussion of the isosinglet axial currents in Sect. X. In the present work
we do not consider an extended version of the model that explicitly includes
vector and axial-vector degrees of freedom \cite{Birse95} (we take $G_{V}=0$,
therefore $g_{A}=1$ and $M_{V}^{2},M_{A}^{2}\rightarrow \infty $).

In order to compute physical quantities we must first determine the full
quark propagator and the full vertices for the vector and axial-vector
currents. All calculations will be done in the leading order of the $1/N_{c}$
expansion, also referred to as the one-quark-loop level or the ladder
approximation. In the nonlocal model the dressed quark propagator, $S(p)$,
with the momentum-dependent quark scalar self-energy (mass),  $M(p)$, is
defined as
\begin{equation}
S^{-1}(p)=\widehat{p}-M(p).  \label{QuarkProp}
\end{equation}%
\begin{figure}[b]
\includegraphics[height=4cm]{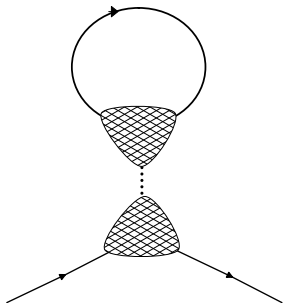}
\caption{{Diagrammatic representation of the quark scalar self-energy of
Eq.~(\protect\ref{SDEq}).}}
\label{w3}
\end{figure}
Note that the considered model involves a constant quark wave-function
renormalization function, $Z(p)=1$. The equation for the quark propagator in
the ladder approximation, also known as the gap equation,
\begin{equation}
M(p)=4iGN_{f}N_{c}f^{2}(p)\int \frac{d^{4}k}{\left( 2\pi \right) ^{4}}%
f^{2}(k)\frac{M(k)}{k^{2}-M^{2}(k)}  \label{SDEq}
\end{equation}%
has the formal solution \cite{Birse95} of the form
\begin{equation}
M(p)=M_{q}f^{2}(p),  \label{Mp}
\end{equation}%
with constant $M_{q}\equiv M(0)$ determined dynamically from Eq.~(\ref{SDEq}%
). The quark self-energy is depicted in Fig. (\ref{w3}). Note that the
functions $f(p)$ are treated non-dynamically, \emph{i.e.} their dependence
on $p$ is fixed, while $M(p)$ is dynamical. Furthermore, the integrals over
the momentum are calculated by transforming the integration variables into
the Euclidean space, ($k^{0}\rightarrow ik_{4},$ $k^{2}\rightarrow -k^{2}$).

\section{Conserved vector and axial-vector currents}

The Noether currents and the corresponding vertices are formally obtained as
functional derivatives of the action (\ref{Lint}) with respect to the
external fields at the zero value of the fields. For our purpose, it is
necessary to construct the quark-current vertices that involve one or two
currents (contact terms). In the presence of the nonlocal interaction the
conserved currents include both local and nonlocal terms. In order to expand
the path-ordered exponent in the external fields, we use the technique
described in \cite{Tern91} (see also \cite{Birse95,ADoLT00}). Briefly, this
method is as follows. First, the Fourier transform of the interaction kernel
in Eq.~(\ref{Lint}) is obtained and expanded in the Taylor series in momemta.
Next, the momentum powers are replaced by the derivatives acting on both the
path-ordered exponent and the quark fields. Finally, the inverse Fourier
transform is performed and summation is carried out again. The relations (%
\ref{MinInt}) and
\begin{equation}
\int d^{4}xF\left( x^{2}\right) e^{-ipx}\int_{y}^{\lambda x+a}dz^{\mu }\
e^{-iqz}=i\lambda \left( 2p+q\lambda \right) ^{\mu }\frac{F\left( p+\lambda
q\right) -F\left( p\right) }{\left( p+\lambda q\right) ^{2}-p^{2}}%
e^{-iqa}+F\left( p\right) \int_{y}^{a}dz^{\mu }\ e^{-iqz},  \label{Rule}
\end{equation}
where $F(z^{2})$ is an arbitrary function, are frequently used in the
procedure described above \footnote{%
We use the same symbol for the function and its four-dimensional Fourier
transform. That should cause no confusion, since one can distinguish the
functions by the notation in their arguments, $x$ or $p$, \textit{etc}.}.
The longitudinal projection of the above relation is
\begin{equation}
q^{\mu }\int d^{4}xF\left( x^{2}\right) e^{-ipx}\int_{y}^{\lambda
x+a}dz^{\mu }\ e^{-iqz}=i\left[ F\left( p+\lambda q\right) e^{-iqa}-F\left(
p\right) e^{-iqy}\right] .
\end{equation}
The algebra needed to obtain the vertices with this method is
straightforward but somewhat tedious, hence below we present only the final
results.

\begin{figure}[]
\includegraphics[width=14cm]{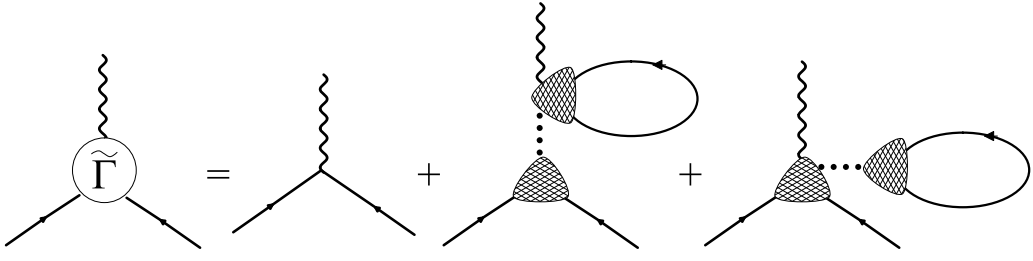}
\caption{{Diagrammatic representation of the bare quark-current vertices (%
\protect\ref{GV}) and (\protect\ref{GAbare}).}}
\label{w5}
\end{figure}

The vector vertex following from the model (\ref{Lint}) is (Fig. \ref{w5})
\begin{equation}
\Gamma _{\mu }^{a}(k,q,k^{\prime }=k+q)=T^{a}\left[ \gamma _{\mu
}-(k+k^{\prime })_{\mu }M^{(1)}(k,k^{\prime })\right] ,  \label{GV}
\end{equation}%
where $M^{(1)}(k,k^{\prime })$ is the finite-difference derivative of the
dynamical quark mass, $q$ is the momentum corresponding to the current, and $%
k$ $(k^{\prime })$ is the incoming (outgoing) momentum of the quark, $
k^{\prime }=k+q$. The finite-difference derivative of an arbitrary function $%
F$ is defined as
\begin{equation}
F^{(1)}(k,k^{\prime })=\frac{F(k^{\prime })-F(k)}{k^{\prime 2}-k^{2}}.
\label{FDD}
\end{equation}%
Thus, with the gauging prescription given by (\ref{Lint}) and (\ref{MinInt}%
), one gets the minimum-coupling vector vertex without extra transverse
pieces. The form of the vertex is the same as the longitudinal vector vertex
resulting from the Pagels-Stokar construction \cite{pagsto79}. The vertex
satisfies the proper Ward-Takahashi identity:%
\begin{equation}
q_{\mu }\Gamma _{\mu }^{a}(k,q,k^{\prime })=S_{F}^{-1}\left( k^{\prime
}\right) T^{a}-T^{a}S_{F}^{-1}\left( k\right) .
\end{equation}%
The vector vertex (\ref{GV}) is free of kinematic singularities and for this
reason was advocated long ago in \cite{pagsto79,BallChiu}. For the case of
the momentum-independent mass, as in local models, the vertex (\ref{GV})
reduces to the usual local form, $\Gamma _{\mu }^{a}=T^{a}\gamma _{\mu }$.

The \emph{bare} axial-vector vertex obtained from the action (\ref{Lint}) by
the differentiation with respect to the fields is given by the formula (%
\emph{cf.} Fig. \ref{w5})
\begin{eqnarray}
\widetilde{\Gamma }_{\mu }^{5a}(k,q,k^{\prime } &=&k+q)=T^{a}\left[ \gamma
_{\mu }-(k+k^{\prime })_{\mu }\frac{\left( \sqrt{M(k^{\prime })}-\sqrt{M(k)}%
\right) ^{2}}{k^{\prime 2}-k^{2}}+\right.  \label{GAbare} \\
&&\left. +\frac{q_{\mu }}{q^{2}}2\sqrt{M(k^{\prime })M(k)}\left[ \frac{G}{%
M_{q}^{2}}J_{AP}(q^{2})-1\right] \right] \gamma _{5},  \nonumber
\end{eqnarray}%
where we have introduced the notation
\begin{eqnarray}
J_{AP}(q^{2}) &=&4N_{c}N_{f}\int \frac{d^{4}l}{\left( 2\pi \right) ^{4}}%
\frac{M\left( l\right) }{D\left( l\right) }\sqrt{M\left( l+q\right) M\left(
l\right) },  \label{JAP} \\
\qquad J_{AP}(q^{2} &\rightarrow &0)=\frac{M_{q}^{2}}{G}-q^{2}J_{AP}^{\prime
}(0)+O\left( Q^{4}\right) ,
\end{eqnarray}%
with%
\begin{equation}
J_{AP}^{\prime }(0)=\frac{N_{c}N_{f}}{32\pi ^{2}}\int du\frac{uM\left(
u\right) \left[ 4M^{\prime }\left( u\right) +2uM^{\prime \prime }\left(
u\right) \right] -u\left( M^{\prime }\left( u\right) \right) ^{2}}{D\left(
u\right) },  \label{J1AP0}
\end{equation}%
where $u=k^2$ and (in the Euclidean space)
\begin{equation}
D\left( k\right) =k^{2}+M^{2}(k).
\end{equation}

\begin{figure}[b]
\includegraphics[height=4cm]{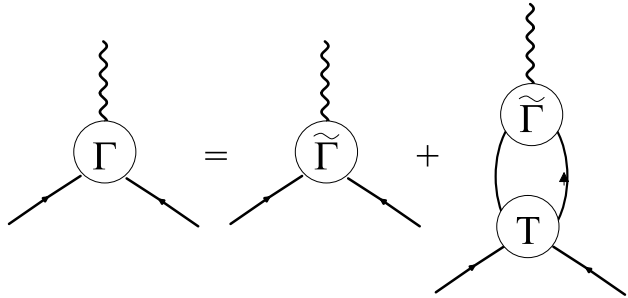}
\caption{{Diagrammatic representation of the full axial-vector vertex
obtained from the bare axial vertex supplied with rescattering process of
Fig. \protect\ref{w4}. In the present model there is no rescattering in the
vector channel, and $\Gamma_V=\tilde\Gamma_V$. }}
\label{w6}
\end{figure}

\begin{figure}[]
\includegraphics[width=14cm]{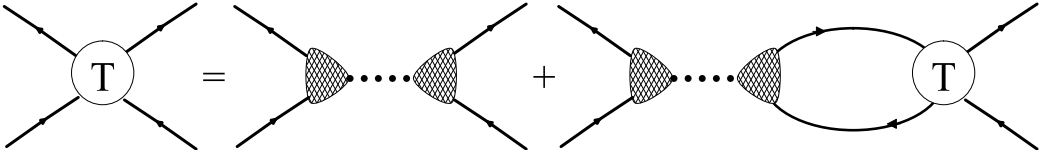}
\caption{{Diagrammatic representation of the quark-quark scattering matrix, $%
T$. }}
\label{w4}
\end{figure}

In Refs.~\cite{HoldLew95,ADoLT00} it was demonstrated that in order to
obtain the full vertex corresponding to the conserved axial-vector current
it is necessary to add the term which contains the pion propagator. The
presence of this term is associated with the well-known pion--axial vector
mixing. The addition of the term with the pion propagator exactly cancels
the third term in Eq.~(\ref{GAbare}), and the full conserved vertex acquires
the form (\emph{cf.} Figs. (\ref{w6}) and (\ref{w4}))
\begin{equation}
\Gamma _{\mu }^{5a}(k,q,k^{\prime }=k+q)=T^{a}\left[ \gamma _{\mu }-q_{\mu }%
\frac{M(k^{\prime })+M(k)}{q^{2}}-(k+k^{\prime }-q\frac{k^{\prime 2}-k^{2}}{%
q^{2}})_{\mu }\frac{\left( \sqrt{M(k^{\prime })}-\sqrt{M(k)}\right) ^{2}}{%
k^{\prime 2}-k^{2}}\right] \gamma _{5}.  \label{GAtot}
\end{equation}%
It satisfies the axial Ward-Takahashi identity,
\begin{equation}
q_{\mu }\Gamma _{\mu }^{5a}(k,q,k^{\prime })=\gamma _{5}S_{F}^{-1}\left(
k_{+}\right) T^{a}+T^{a}S_{F}^{-1}\left( k_{-}\right) \gamma _{5}.
\label{AxWTI}
\end{equation}%
The axial-vector vertex has a kinematic pole at $q^{2}=0$, a property that
follows from the spontaneous breaking of the chiral symmetry in the limit of
massless $u$ and $d$ quarks. Evidently, this pole corresponds to the
massless Goldstone pion.

We also need the vertices that couple two currents to the quark (\emph{cf.}
Fig. \ref{w2}). In this regard it is convenient to introduce the notation
\begin{equation}
G_{\mu }^{a}\left( k,q\right) =iT^{a}\left( 2k+q\right) _{\mu}f^{(1)}(k,k+q),
\end{equation}%
and%
\begin{eqnarray}
G_{\mu \nu }^{ab}\left( k,q,q^{\prime },k^{\prime }\right) &=&-f\left(
k^{\prime }\right) \left\{ T^{a}T^{b}\left[ g_{\mu \nu
}f^{(1)}(k,k+q+q^{\prime })+\right. \right. \\
&+&\left. \left. \left[ 2\left( k+q^{\prime }\right) +q\right] _{\mu }\left(
2k+q^{\prime }\right) _{\nu }f^{(2)}\left( k,k+q^{\prime },k+q+q^{\prime
}\right) \right] +\left[ \left( q,a,\mu \right) \longleftrightarrow \left(
q^{\prime },b,\nu \right) \right] \right\} ,  \nonumber
\end{eqnarray}%
where the second finite-difference derivative is defined by%
\begin{equation}
F^{(2)}\left( k,k^{\prime },k^{\prime \prime }\right) =\frac{%
F^{(1)}(k,k^{\prime \prime })-F^{(1)}(k,k^{\prime })}{k^{\prime \prime
2}-k^{\prime 2}}.
\end{equation}%
Further, we need to introduce
\begin{eqnarray}
F_{\mu }^{(\pm )a}\left( k,q\right) &=&G_{\mu }^{a}\left( k,q\right) \pm
G_{\mu }^{a}\left( k-q,q\right) , \\
F_{\mu \nu }^{(\pm )ab}\left( k,q,q^{\prime },k^{\prime }\right) &=&G_{\mu
\nu }^{ab}\left( k,q,q^{\prime },k^{\prime }\right) +G_{\mu \nu }^{ab}\left(
k^{\prime }-q-q^{\prime },q,q^{\prime },k\right)  \nonumber \\
&\pm &G_{\mu }^{a}\left( k,q\right) G_{\nu }^{b}\left( k^{\prime }-q^{\prime
},q^{\prime }\right) \pm G_{\mu }^{a}\left( k^{\prime }-q,q\right) G_{\nu
}^{b}\left( k,q^{\prime }\right) .
\end{eqnarray}%
With the above definitions one gets for the $VV$ contact term%
\begin{eqnarray}
\Gamma _{\mu \nu }^{ab}(k,q,q^{\prime },k^{\prime }=k+q+q^{\prime })
&=&M_{q}F_{\mu \nu }^{(+)ab}\left( k,q,q^{\prime },k^{\prime }=k+q+q^{\prime
}\right)  \label{GVV} \\
&+&f(k)f(k^{\prime })G\int \frac{d^{4}l}{\left( 2\pi \right) ^{4}}Tr\left[
S(l)F_{\mu \nu }^{(+)ab}\left( l,q,q^{\prime },l\right) \right] .  \nonumber
\end{eqnarray}
For the $AA$ contact term one finds
\begin{equation}
\Gamma _{\mu \nu }^{5ab}(k,q,q^{\prime },k^{\prime }=k+q+q^{\prime })=\Gamma
_{(1)\mu \nu }^{5ab}(k,q,q^{\prime },k^{\prime }=k+q+q^{\prime })+\Delta
\Gamma _{\mu \nu }^{5ab}(k,q,q^{\prime },k^{\prime }=k+q+q^{\prime }),
\label{GAA}
\end{equation}%
where%
\begin{eqnarray}
\Gamma _{(1)\mu \nu }^{5ab}(k,q,q^{\prime },k^{\prime }=k+q+q^{\prime })
&=&M_{q}F_{\mu \nu }^{(-)ab}\left( k,q,q^{\prime },k^{\prime }=k+q+q^{\prime
}\right)  \label{GAA1} \\
&+&f(k)f(k^{\prime })G\int \frac{d^{4}l}{\left( 2\pi \right) ^{4}}Tr\left[
S(l)F_{\mu \nu }^{(-)ab}\left( l,q,q^{\prime },l\right) \right] .
\end{eqnarray}%
An additional contribution appears for the $AA$ iso-triplet contact term
\begin{eqnarray}
\Delta \Gamma _{\mu \nu }^{5ab}(k,q,q^{\prime },k^{\prime } &=&k+q+q^{\prime
})=-G\left[ \tau ^{c}G_{\nu }^{b}\left( k,q^{\prime }\right) -G_{\nu
}^{b}\left( k-q^{\prime },q^{\prime }\right) \tau ^{c}\right]  \label{GAA2}
\\
&\times &\left[ \int \frac{d^{4}l}{\left( 2\pi \right) ^{4}}Tr\left[
S(l)\tau ^{c}F_{\mu }^{(-)a}\left( l,q\right) \right] \right] +\left[ \left(
q,\mu ,a\right) \longleftrightarrow \left( q^{\prime },\nu ,b\right) \right]
.  \nonumber
\end{eqnarray}%
In the above expressions $Tr$ denotes the trace over flavor, color, and
Dirac indices.

\begin{figure}[b]
\includegraphics[width=6cm]{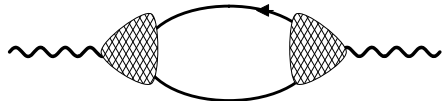}
\caption{The polarization operator of Eq. (\protect\ref{Ppp})}
\label{w7}
\end{figure}

\begin{figure}[tbp]
\includegraphics[width=5cm]{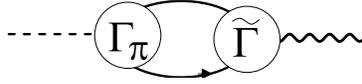}
\caption{The correlator of the bare axial current vertex (\protect\ref%
{GAbare}) and the pion vertex (\protect\ref{Gpi}).}
\label{w8}
\end{figure}

In the following we also need to introduce the polarization operator in the
pseudoscalar channel (\emph{cf.} Fig. \ref{w7}),
\begin{equation}
J_{PP}(q^{2})\delta _{ab}=-\frac{i}{M_{q}^{2}}\int \frac{d^{4}k}{\left( 2\pi
\right) ^{4}}M\left( k\right) M\left( k+q\right) Tr\left[ S(k)\gamma
_{5}\tau ^{a}S\left( k+q\right) \gamma _{5}\tau ^{b}\right]  \label{Ppp}
\end{equation}%
and the correlator of the axial current vertex (\ref{GAbare}) and the pion
vertex (\emph{cf.} Fig. \ref{w8})
\begin{equation}
\Gamma _{\pi }^{a}\left( k,k^{\prime }\right) =ig_{\pi }\gamma
_{5}f(k)f(k^{\prime })\tau ^{a},  \label{Gpi}
\end{equation}%
defined by%
\begin{equation}
J_{\pi A}\left( q^{2}\right) \delta _{ab}=\frac{q_{\mu }}{q^{2}}\int \frac{%
d^{4}k}{\left( 2\pi \right) ^{4}}Tr\left[ S(k)\widetilde{\Gamma }_{\mu
}^{5a}(k,q,k+q)S\left( k+q\right) \Gamma _{\pi }^{a}\left( k+q,k\right) %
\right] .  \label{PpiA}
\end{equation}

Through the use of the gap equation (\ref{SDEq}) and the expression for the
pion decay constant, $f_{\pi }$, given by \cite{DP86,Birse95}
\begin{equation}
f_{\pi }^{2}=\frac{N_{c}}{4\pi ^{2}}\int\limits_{0}^{\infty }du\ u\frac{%
M(u)^{2}-uM(u)M^{\prime }(u)+u^{2}M^{\prime }(u)^{2}}{D^{2}\left( u\right) },
\label{Fpi2_M}
\end{equation}%
these correlators have the following expansion at zero momentum:
\begin{equation}
J_{PP}(q^{2})=\frac{1}{G}+\frac{f_{\pi }^{2}}{M_{q}^{2}}q^{2}+O\left(
q^{4}\right) ,\qquad J_{\pi A}\left( q^{2}\right) =f_{\pi }^{2}+O\left(
q^{2}\right) .  \label{PpiA0}
\end{equation}%
In (\ref{Fpi2_M}) we have used the notation $u=k^{2}$ and $M^{\prime
}(u)=dM(u)/du$. In Eq.~(\ref{Gpi}) the quark-pion coupling, $g_{\pi }^{2}=%
\left[ J_{PP}^{\prime }(0)\right] ^{-1}$, and the pion decay constant, $%
f_{\pi }$, are connected by the Goldberger-Treiman relation,
\begin{equation}
g_{\pi }=\frac{M_{q}}{f_{\pi }},  \label{GTR}
\end{equation}%
which is verified to be valid in the nonlocal model \cite{Birse95}, as
required by the chiral symmetry.

\section{Current-current correlators (transverse parts)}

\begin{figure}[b]
\includegraphics[width=14.5cm]{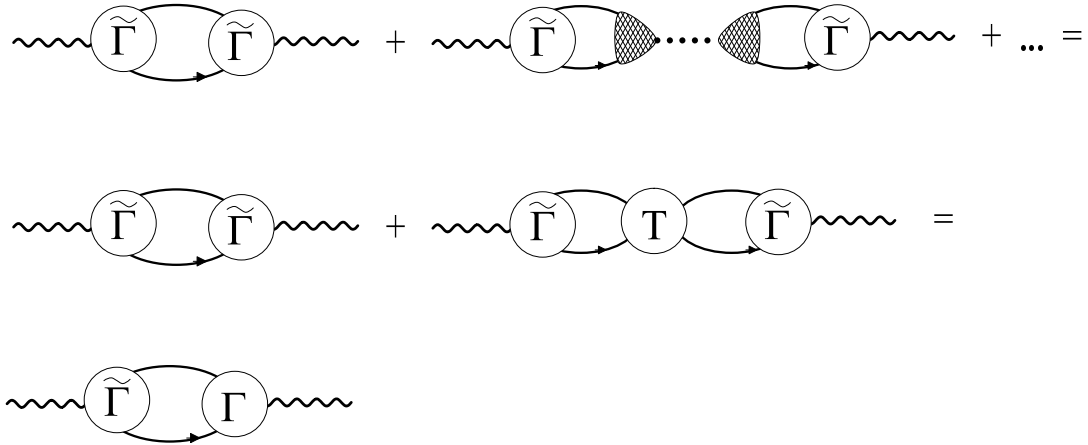}
\caption{The resummation of the quark rescattering in the current-current
correlator in the axial-vector channel. }
\label{w9}
\end{figure}

\begin{figure}[]
\includegraphics[width=12cm]{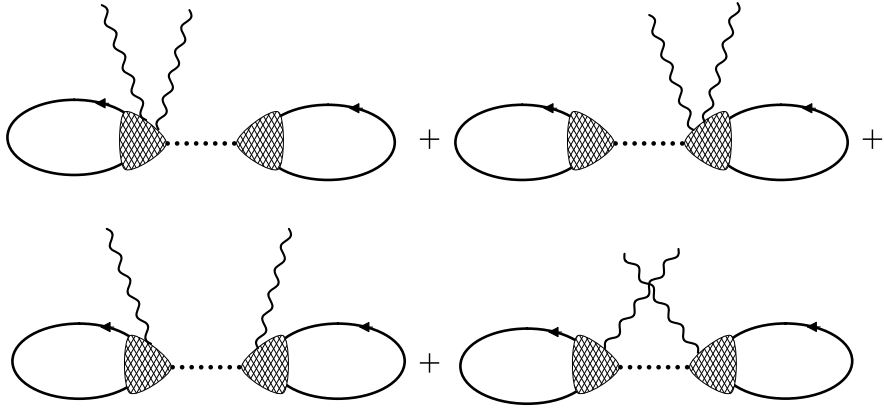}
\caption{The contact terms in the current-current correlators. }
\label{w10}
\end{figure}

Our goal is to obtain the nonperturbative parts of the current-current
correlators from the effective model and to compare them with the existing $%
\tau $ decay data. The current-current correlators may be represented as a
sum of two terms, dispersive (Fig. \ref{w9}) and contact (Fig. \ref{w10}),
namely
\begin{eqnarray}
-Q^{2}\Pi _{\mu \nu }^{J}\left( Q^{2}\right) &=&K_{\mu \nu }^{J}\left(
Q^{2}\right) +S_{\mu \nu }^{J}\left( Q^{2}\right) ,  \label{Q2P} \\
K_{\mu \nu }^{V}\left( Q^{2}\right) &=&\int \frac{d^{4}k}{\left( 2\pi
\right) ^{4}}Tr\left[ \Gamma _{\mu }^{V}\left( k,Q,k+Q\right) S\left(
k+Q\right) \Gamma _{\nu }^{V}\left( k+Q,-Q,k\right) S\left( k\right) \right]
,  \label{KV} \\
K_{\mu \nu }^{A}\left( Q^{2}\right) &=&\int \frac{d^{4}k}{\left( 2\pi
\right) ^{4}}Tr\left[ \Gamma _{\mu }^{A}\left( k,Q,k+Q\right) S\left(
k+Q\right) \widetilde{\Gamma }_{\nu }^{A}\left( k+Q,-Q,k\right) S\left(
k\right) \right] ,  \label{KA} \\
S_{\mu \nu }^{J}\left( Q^{2}\right) &=&2M_{q}\int \frac{d^{4}k}{\left( 2\pi
\right) ^{4}}Tr\left[ S\left( k\right) \Gamma _{\mu \nu }^{J}\left(
k,Q,-Q,k+Q\right) \right] .  \label{SJ}
\end{eqnarray}%
The vertices $\Gamma _{\mu }^{J}\left( k,q,k^{\prime }\right) $ are given in
Eqs. (\ref{GV},\ref{GAtot}), $\Gamma _{\mu \nu }^{J}\left( k,q,q^{\prime
},k^{\prime }\right) $ in Eq. (\ref{GVV},\ref{GAA}), and $\widetilde{\Gamma }%
_{\mu \nu }^{A}\left( k,q,q^{\prime },k^{\prime }\right) $ in Eq. (\ref%
{GAbare}). The difference in the definitions of $K_{\mu \nu }\left(
Q^{2}\right) $ in (\ref{KV}) and (\ref{KA}) results from the necessity of
taking into account the rescattering diagrams in the axial channel of the
pseudoscalar ($\pi $ or $\eta ^{\prime }$) mesons (Fig. \ref{w9}). The
correlators (\ref{Q2P}) are defined in such a way that the perturbative
contributions are subtracted,
\begin{equation}
\Pi _{np}\left( Q^{2}\right) =\Pi _{tot}\left( Q^{2}\right) -\Pi
_{pert}\left( Q^{2}\right) .
\end{equation}%
The perturbative part is obtained from the non-perturbative part by simply
setting the dynamical quark mass $M(k)$ to zero. We extract the longitudinal and transverse
parts of the correlators through the use of the projectors%
\begin{equation}
P_{\mu \nu }^{L}=\frac{q^{\mu }q^{\nu }}{q^{2}},\qquad P_{\mu \nu }^{T}=%
\frac{1}{3}\left( g_{\mu \nu }-\frac{q^{\mu }q^{\nu }}{q^{2}}\right) .
\label{LTprojectors}
\end{equation}

We first consider the transverse part of the $V$ correlator, with the result
\begin{eqnarray}
K_{T}^{V}\left( Q^{2}\right) &=&2N_{c}\int \frac{d^{4}k}{\left( 2\pi \right)
^{4}}\frac{1}{D_{+}D_{-}}\left\{ M_{+}M_{-}+\left[ k_{+}k_{-}-\frac{2}{3}%
k_{\perp }^{2}\right] _{np}\right. \\
&+&\left. \frac{4}{3}k_{\perp }^{2}\left[ \left( \frac{M_{+}-M_{-}}{%
k_{+}^{2}-k_{-}^{2}}\right) ^{2}\left( k_{+}k_{-}-M_{+}M_{-}\right) -\frac{%
M_{+}^{2}-M_{-}^{2}}{k_{+}^{2}-k_{-}^{2}}\right] \right\} , \nonumber\\
S_{T}^{V}\left( Q^{2}\right) &=&4N_{c}\int \frac{d^{4}k}{\left( 2\pi \right)
^{4}}\frac{M\left( k\right) }{D\left( k\right) }\left\{ M^{\prime }\left(
k\right) +\frac{4}{3}\frac{k_{\perp }^{2}}{k^{2}-\left( k+Q\right) ^{2}}%
\left( M^{\prime }\left( k\right) -\frac{M\left( k+Q\right) -M\left(
k\right) }{\left( k+Q\right) ^{2}-k^{2}}\right) \right\} ,
\end{eqnarray}%
where the notations $k_{\perp }^{\mu }=k^{\mu }-Q^{\mu }(k Q)/Q^{2}$, $%
k_{\pm }=k\pm Q/2$,
\begin{equation}
M_\pm = M(k_\pm),\ \ \ \ \ D_\pm = D(k_\pm)
\end{equation}
have been introduced. The subtraction of the perturbative part amounts to
the replacement
\begin{equation}
\frac{1}{D_{+}D_{-}} \left[ k_{+}k_{-}-\frac{2}{3}k_{\perp }^{2}\right]
_{np}\Longrightarrow \left[ k_{+}k_{-}-\frac{2}{3}k_{\perp }^{2}\right] %
\left[ \frac{1}{D_{+}D_{-}}-\frac{1}{k_{+}^{2}k_{-}^{2}}\right] .
\end{equation}%
Further, we take the nonsinglet transverse projection of the $A$ correlator
and obtain
\begin{eqnarray}
K_{T}^{A}\left( Q^{2}\right) &=&2N_{c}\int \frac{d^{4}k}{\left( 2\pi \right)
^{4}}\frac{1}{D_{+}D_{-}}\left\{ -M_{+}M_{-}+\left[ k_{+}k_{-}-\frac{2}{3}%
k_{\perp }^{2}\right] _{np}+\right.   \\
&+&\left. \frac{4}{3}k_{\perp }^{2}\left[ \frac{\left( \sqrt{M_{+}}-\sqrt{%
M_{-}}\right) ^{4}}{\left( k_{+}^{2}-k_{-}^{2}\right) ^{2}}\left(
k_{+}k_{-}+M_{+}M_{-}\right) -\frac{\left( M_{+}-M_{-}\right) \left( \sqrt{%
M_{+}}-\sqrt{M_{-}}\right) ^{2}}{k_{+}^{2}-k_{-}^{2}}\right] \right\} , \nonumber\\
S_{T}^{A}\left( Q^{2}\right) &=&S_{T}^{V}\left( Q^{2}\right) -4N_{c}\int
\frac{d^{4}k}{\left( 2\pi \right) ^{4}}\frac{M\left( k\right) }{D\left(
k\right) }\frac{8}{3}k_{\perp }^{2}\frac{\left( \sqrt{M\left( k+Q\right) }-%
\sqrt{M\left( k\right) }\right) ^{2}}{\left[ \left( k+Q\right) ^{2}-k^{2}%
\right] ^{2}}.
\end{eqnarray}

Let us consider the difference of the $V$ and $A$ correlators, where a
number of cancellations takes place and the final result is quite simple,
\begin{eqnarray}
-Q^{2}\Pi ^{V-A}\left( Q^{2}\right) &=&4N_{c}\int \frac{d^{4}k}{\left( 2\pi
\right) ^{4}}\frac{1}{D\left( k_{+}^{2}\right) D\left( k_{-}^{2}\right) }%
\left\{ M_{+}M_{-}+\frac{4}{3}k_{\perp }^{2}\left[ -\sqrt{M_{+}M_{-}}\frac{%
M_{+}-M_{-}}{k_{+}^{2}-k_{-}^{2}}\right. \right.  \label{VmAmodel} \\
&+&\left. \left. \frac{\left( \sqrt{M_{+}}-\sqrt{M_{-}}\right) ^{2}}{\left(
k_{+}^{2}-k_{-}^{2}\right) ^{2}}\left( \sqrt{M_{+}}k_{+}+\sqrt{M_{-}}%
k_{-}\right) ^{2}\right] \right\} .  \nonumber
\end{eqnarray}
One may explicitly verify that the integrand of the above expression is
indeed positive-definite, irrespectively of the choice of the mass function $%
M(p)$. Thus the Witten inequality (\ref{WittIneuq}) is indeed fulfilled.

At $Q^{2}=0$ one gets the results consistent with the first Weinberg sum
rule,
\begin{equation}
-Q^{2}\Pi ^{V,T}\left( Q^{2}=0\right) =0,\qquad -Q^{2}\Pi ^{A,T}\left(
Q^{2}=0\right) =-f_{\pi }^{2},\qquad -Q^{2}\Pi ^{V-A,T}\left( Q^{2}=0\right)
=f_{\pi }^{2},  \label{JT0}
\end{equation}%
where the explicit definition of the pion decay constant (\ref{Fpi2_M}) is
used. This serves as a useful algebraic check.

\section{Model parameters}

The parameters of the model are fixed in a way typical for effective
low-energy quark models. We demand that the pion decay constant $f_{\pi }$, (%
\ref{Fpi2_M}), and the quark condensate (for a single flavor), $<\bar{q}q>$,
given by
\begin{equation}
\langle \bar{q}q\rangle =-\frac{N_{c}}{4\pi ^{2}}\int du\ u\frac{M(u)}{%
D\left( u\right) },  \label{QQcond}
\end{equation}%
acquire their physical values. For simplicity, we take profile for the
dynamical quark mass in a Gaussian form%
\begin{equation}
M(u)=M_{q}\exp \left( -2u/\Lambda ^{2}\right) .
\end{equation}%
With the model parameters%
\begin{equation}
M_{q}=0.3~\mathrm{GeV,}\qquad \Lambda =1.085~\mathrm{GeV,}
\end{equation}%
one obtains
\begin{equation}
f_{\pi }=93~\mathrm{MeV,}\qquad \langle \bar{q}q\rangle =-\left( 224~\mathrm{%
MeV}\right) ^{3},
\end{equation}%
where the quark condensate is supposed to be normalized at the scale of
a few hundred MeV.

\section{Large-$Q^{2}$ expansion}

At large $Q^{2}$ one finds the following asymptotic expansion for the
difference and sum of the correlation functions in the inverse powers of $%
Q^{2}$ (we do not display here the exponentially-suppressed terms coming
from powers of the dynamical quark-mass form factor):
\begin{eqnarray}
\left. -Q^{2}\Pi ^{V-A}\left( Q^{2}\right) \right\vert _{Q^{2}\rightarrow
\infty } &=&\frac{2}{Q^{4}}\frac{N_{c}}{4\pi ^{2}}\int du\frac{%
u^{2}M^{2}\left( u\right) }{D\left( u\right) }+\mathcal{O}\left( \frac{1}{%
Q^{6}}\right) ,  \label{PVmAlarge} \\
-Q^{2}\Pi _{T}^{V+A}\left( Q^{2}\rightarrow \infty \right) &=&2\frac{N_{c}}{%
4\pi ^{2}}\int du\frac{u}{D\left( u\right) }\left[ M\left( u\right)
M^{\prime }\left( u\right) \left( 1-\frac{u}{Q^{2}}-\frac{2}{3}\frac{u^{2}}{%
Q^{4}}\right) -\frac{7}{6}\frac{uM^{2}\left( u\right) }{Q^{4}}\right]
+\mathcal{O}\left( \frac{1}{Q^{6}}\right) .\label{PVpAlarge}
\end{eqnarray}

The effective model considered here is designed to describe low energy
physics. At high energies it is certainly not expected to reproduce all the
details of the asymptotic standard operator product expansion of QCD. On
other hand, it is possible that the OPE works well only at very short
distances while the effective model is applicable at large and intermediate
distances. With this hope in mind we proceed to analyzing the large-$Q^{2}$
expansions of the correlators, comparing them numerically to the OPE
results, and trying to match the two approaches. It is important to note
that the power corrections in the expansions (\ref{PVmAlarge}) and (\ref%
{PVpAlarge}) have the same inverse powers of $Q^{2}$ as the OPE.

We may now compare the expansion of the model correlators to the OPE , Eqs. (%
\ref{PVmA},\ref{PVpA}). In Eq. (\ref{PVmAlarge}) the formally leading $d=4$
term absent in the chiral limit in accordance with the second Weinberg sum
rule and the OPE QCD. The second term in Eq.~(\ref{PVmAlarge}) (and the last
term in Eq.~(\ref{PVpAlarge})) is proportional to the derivative of the
gluon condensate, and via equations of motion it reduces to the four-quark
condensate term appearing in the OPE, Eqs.~(\ref{PVmA},\ref{PVpA}). Let us
compare the numerical estimates for the local $d=6$ terms obtained from the
QCD sum rules and from the nonlocal chiral quark model, labeled as N$\chi $%
QM:
\begin{eqnarray}
\left[ O_{6}^{V-A}\right] ^{\mathrm{QCDsr}} &\approx &-(1\div 2)\cdot 10^{-3}%
\mathrm{~GeV}^{6},\qquad  \label{OVmA6} \\
\left[ O_{6}^{V-A}\right] ^{\mathrm{OPE\tau }} &=&-(3.4\pm 1.1)\cdot 10^{-3}%
\mathrm{~GeV}^{6},\qquad  \nonumber \\
\left[ O_{6}^{V-A}\right] ^{N\chi QM} &=&-1.1\cdot 10^{-3}\mathrm{~GeV}^{6}.
\nonumber
\end{eqnarray}%
The first estimate is found on the basis of low energy theorems and QCD sum
rules \cite{SVZ79}, while the second estimate is made with the help of the $%
\tau $-decay data \cite{ioffe}. The result of the present model is closer to
the standard estimate obtained from the low-energy phenomenology. Similar
features of the short-distance behavior of the correlators were found in the
instanton model \cite{SHURYAK}.

In the $V+A$ correlator (\ref{PVpAlarge}) the short-distance expansion
contains, in addition to the contributions coming from the local operators,
the unconventional terms originating from the nonlocal operators of
dimension $d=2,4$ and $6$ (the first terms in Eq.~(\ref{PVpAlarge})). This
kind of unconventional terms has recently attracted attention due to the
revision of the standard OPE \cite{CNZ}, as well as the lattice findings,
where the unconventional power corrections in the vector correlators were
reported \cite{Boucaud:2001st}. The appearance of this correction is usually
related to the existence of the lowest $d=2$ condensate $\left\langle \left(
A_{\mu }^{a}\right) ^{2}\right\rangle $, which is due to an apparent gauge
non-invariance, absent in the standard OPE. However, in a series of papers (%
\cite{Stodolsky:2002st,Dudal:2003vv}, and references therein) it was argued
that it is possible to define the nonlocal operator with the lowest
dimension in a gauge-invariant way. This situation is very similar to the
famous spin-crisis problem (\emph{cf.} \cite{Dorokhov:ym}). Analogously, in
that case there is no twist-two gluonic operator that may contribute to the
singlet axial current matrix element, yet, it is possible to construct the
matrix element from nonlocal operators \cite{Jaffe:1989jz}. We thus see that
our effective nonlocal model shares these unusual effects generated by the
internal nonlocalities of the quark interaction. Furthermore, the
lowest-dimension power corrections are naturally present in the approaches
similar to the analytic perturbative theory \cite{DVSol}. In that case in
order to compensate the effects of the ghost pole in the strong coupling
constant, the $d=2$ power term is added. As discussed in Ref. \cite%
{Narison:2001ix}, the justification of the appearance of the unconventional
power corrections at the same time means that the standard OPE is valid only
at very large momenta.

We also wish to comment that in the model expansion of the $V+A$ correlator
there are no explicit terms with the gluon condensate of dimension $d=4$.
The appearance of this term in the nonlocal chiral quark model would
corresponded to the local matrix element%
\[
\frac{N_{c}}{4\pi ^{2}}\int du\frac{uM^{2}\left( u\right) }{D\left( u\right)
},
\]%
that is related to the gluon condensate through the gap equation (\ref{SDEq}%
) \cite{DP85}. However, the coefficient of this term is equal to zero. This
is due to the simple form of the quark propagator (\ref{QuarkProp}), which
does not allow gluonic correlations between different quark lines. The
similar situation occurs in the QCD sum rules calculations (in the fixed
point gauge), where nonzero contribution comes from the diagram with quark
lines correlated by soft gluon exchange. These (numerically small)
correlation terms may be reconstructed in the effective model by introducing
the gluonic field in the effective action (\ref{Lint}) by gauging kinetic
and interaction terms (see also \emph{cf.} \cite{Bijnens}). From other hand
the $d=4$ term appears in (\ref{PVpAlarge}) as a nonlocal matrix element.

We end the discussion of the short-range behavior of the correlators by
giving the numerical estimates of the additional terms appearing in Eq.~(\ref%
{PVpAlarge}):
\begin{eqnarray}
\left[ O_{2}^{V+A}\right] _{\mathrm{nonloc}}^{N\chi QM} &=&-\frac{N_{c}}{%
2\pi ^{2}}\int du\frac{u}{D\left( u\right) }M\left( u\right) M^{\prime
}\left( u\right) =5.0\cdot 10^{-3}\mathrm{~GeV}^{2}, \\
\left[ O_{4}^{V+A}\right] _{\mathrm{nonloc}}^{N\chi QM} &=&\frac{N_{c}}{2\pi
^{2}}\int du\frac{u^{2}}{D\left( u\right) }M\left( u\right) M^{\prime
}\left( u\right) =-1.8\cdot 10^{-3}\mathrm{~GeV}^{4},  \nonumber \\
\left[ O_{6}^{V+A}\right] _{\mathrm{nonloc}}^{N\chi QM} &=&\frac{4}{3}\frac{%
N_{c}}{4\pi ^{2}}\int du\frac{u^{3}}{D\left( u\right) }M\left( u\right)
M^{\prime }\left( u\right) =-7.6\cdot 10^{-4}\mathrm{~GeV}^{6}.  \nonumber
\end{eqnarray}%
The sum of these terms, taken in the interval of momenta $Q^{2}\sim \left(
1\div 2\right) $ GeV$^{2}$ where the model large-$Q^{2}$ expansion is
expected to be valid, agrees reasonably well with the coefficient of the
leading power correction in Eq.~(\ref{PVpA})
\begin{equation}
\left[ O_{2}^{V+A}\right] ^{\mathrm{QCDsr}}=3.0\cdot 10^{-3}\mathrm{~GeV}%
^{2},
\end{equation}%
where we have taken the estimate $\left( \alpha _{s}/\pi \right) \lambda
^{2}=-0.12\mathrm{~GeV}^{2}$ from Ref.~\cite{Narison:2001ix}.

Through the use of the factorization hypothesis (\ref{O6factor}) it is predicted that the chirality
flip matrix element $O_{6}^{V-A}$ is strongly enhanced in absolute value
over the chirality nonflip one $O_{6}^{V+A}$%
\begin{equation}
\left[ O_{6}^{V-A}/O_{6}^{V+A}\right] ^{\mathrm{factor}}=-4.5.  \label{O6factorRatio}
\end{equation}%
In the nonlocal chiral quark model the chirality flip matrix element $%
O_{6}^{V-A}$ is given by the local matrix element, but the chirality nonflip
one $O_{6}^{V+A}$ is a mixture of the local and nonlocal matrix elements
which transform to each other by integration by parts. We find that their
ratio
\begin{equation}
\left[ O_{6}^{V-A}/O_{6}^{V+A}\right] ^{N\chi QM}\approx -3.2
\label{O6modelRatio}
\end{equation}%
has the same tendency as predicted in (\ref{O6factorRatio}). It happens
due to partial compensation of contributions of the local and nonlocal
matrix elements into $O_{6}^{V+A}.$

\section{Low-energy observables and the ALEPH data}

Let us now consider the low-energy region where the effective model (\ref%
{Lint}) should be fully predictive. From (\ref{VmAmodel}) and the DGMLY sum
rule (\ref{DGMLY}) we estimate the electromagnetic pion mass difference to
be
\begin{equation}
\left[ m_{\pi ^{\pm }}-m_{\pi ^{0}}\right] _{N\chi QM}=4.2\mathrm{~MeV},
\label{dMmodel}
\end{equation}%
which is in remarkable agreement with the experimental value (after
subtracting the $m_{d}-m_{u}$ effect) \cite{GasLeut85}
\begin{equation}
\left[ m_{\pi ^{\pm }}-m_{\pi ^{0}}\right] _{\mathrm{exp}}=4.43\pm 0.03%
\mathrm{~MeV}.
\end{equation}

It is interesting to estimate the electric polarizability of the charged
pions, \cite{Petrun,DorVoHuKle97}. With the help of the DMO sum rule \footnote{
In $\chi $PT in the one-loop approximation the right-hand-side of the DMO
sum rule is expressed through the low-energy constant $L_{10}$. The
extraction of this constant from the experiment was considered in \cite%
{Davier:1998dz}.}, as done by Gerasimov in \cite{Geras79}, we find
\begin{equation}
\alpha^E _{\pi ^{\pm }}=\frac{\alpha }{m_{\pi }}\left[ \frac{\left\langle
r_{\pi }^{2}\right\rangle }{3}-\frac{I_{DMO}}{f_{\pi }^{2}}\right] ,
\label{PiPolariz}
\end{equation}%
where $I_{DMO}$ is the left-hand side of the DMO sum rule (\ref{DMO})%
\begin{equation}
I_{DMO}\left( s_{0}\right) =\frac{1}{4\pi ^{2}}\int_{0}^{s_{0}}\frac{ds}{s}%
\left[ v_{1}\left( s\right) -a_{1}\left( s\right) \right] .  \label{Idmo}
\end{equation}%
Equation (\ref{PiPolariz}) can be interpreted as a sum of the center-of-mass
recoil contribution and the intrinsic pion polarizability. In \cite{Geras79}
it was demonstrated that in model calculations there occurs a delicate
cancellation between the two contributions of Eq.~(\ref{PiPolariz}). This
requires the calculation of both terms consistently at the same level of
approximations.

With the experimental value for the pion mean squared radius \cite{NA7} and
the value of the $I_{DMO}$ integral estimated from the ALEPH and OPAL data
\cite{OPAL}
\begin{equation}
\left[ I_{DMO}\left( m_{\tau }^{2}\right) \right] _{\mathrm{exp}}=\left(
26.3\pm 1.8\right) \cdot 10^{-3}
\end{equation}%
one gets from Eq. (\ref{PiPolariz}) the result \cite{OPAL}
\begin{equation}
\left[ \alpha^E _{\pi ^{\pm }}\right] _{\mathrm{exp}}=\left( 2.71\pm
0.88\right) \cdot 10^{-4}\mathrm{~fm}^{3}.  \label{PiPolarizOPAL}
\end{equation}

From Eq.~(\ref{DMO}) also follows the relation obtained by Terentyev \cite%
{Teren72}, which relates the pion polarizability and the pion axial-vector
form factor,
\begin{equation}
\alpha^E _{\pi ^{\pm }}=\frac{\alpha F_{A}}{m_{\pi }f_{\pi }^{2}}.
\label{PiPolarizFA}
\end{equation}%
The last relation, used with the known values for $F_{A}$, yields
\begin{equation}
\alpha^E _{\pi ^{\pm }}=\left( 2.69\pm 0.37\right) \cdot 10^{-4}\mathrm{~fm}%
^{3},
\end{equation}%
which is very close to (\ref{PiPolarizOPAL}).

Let us estimate the electric polarizability of the charged pions within the
nonlocal chiral quark model. By calculating the derivative of $\Pi
^{V-A}\left( Q^{2}\right) $ at zero momentum we estimate the left hand side
of the DMO sum rule as
\begin{equation}
\left[ I_{DMO}\left( s_{0}\rightarrow \infty \right) \right] _{N\chi
QM}=18.2\cdot 10^{-3}.  \label{L10m}
\end{equation}%
The value of the pion charge radius squared,
\begin{equation}
\left[ \left\langle r_{\pi }^{2}\right\rangle \right] _{N\chi QM}=0.33\quad
\mathrm{fm}^{2},  \label{r2pim}
\end{equation}%
obtained in our model from the derivative of the charged pion form factor, is close to
its limit of the local chiral model, found by Gerasimov long ago \cite%
{Geras79},
\begin{equation}
\left[ \left\langle r_{\pi }^{2}\right\rangle \right] _{\chi PT}=\frac{N_{c}%
}{4\pi ^{2}f_{\pi }^{2}}=0.34\quad \mathrm{fm}^{2}.  \label{r2pi}
\end{equation}%

The model predictions for $\left\langle r_{\pi }^{2}\right\rangle $ and $%
I_{DMO}$ are somewhat smaller than the experimental values given above.
The reason for this discrepancy
may be attributed to vector meson degrees of freedom, neglected in our treatment,
and to pion loops absent in the large-$N_{c}$ limit.
However, these unconsidered contributions are essentially canceled in the
combination (\ref{Peuclid}) defining the electric pion polarizability (see,
{\it e.g.} \cite{Klevansky:1997dk} for discussions). From (\ref{PiPolariz}) we
find with values given in Eqs. (\ref{L10m}) and (\ref{r2pim}) the value
\begin{equation}
\left[ \alpha^E _{\pi ^{\pm }}\right] _{N\chi QM}=2.9\cdot 10^{-4}\mathrm{fm}%
^{3},  \label{AlMod}
\end{equation}%
which is close to experimental number (\ref{PiPolarizOPAL}) and also to the
prediction of the chiral perturbation theory at the one-loop level \cite%
{burgi},
\begin{equation}
\left[ \alpha^E _{\pi ^{\pm }}\right] _{\chi \mathrm{PT}}=2.7\cdot 10^{-4}%
\mathrm{fm}^{3}.
\end{equation}%
Let us note also that (\ref{AlMod}) is a factor of 2 smaller from the
estimates obtained in a local chiral quark model \cite{KlevPolariz99}, $%
\alpha^E _{\pi ^{\pm }}=5.8\cdot 10^{-4}\mathrm{~fm}^{3}$. We thus see that
the model prediction for the pion polarizability, Eq.~(\ref{AlMod}), is in a
very reasonable agreement with the experimental data.

\begin{figure}[]
\includegraphics[width=10cm]{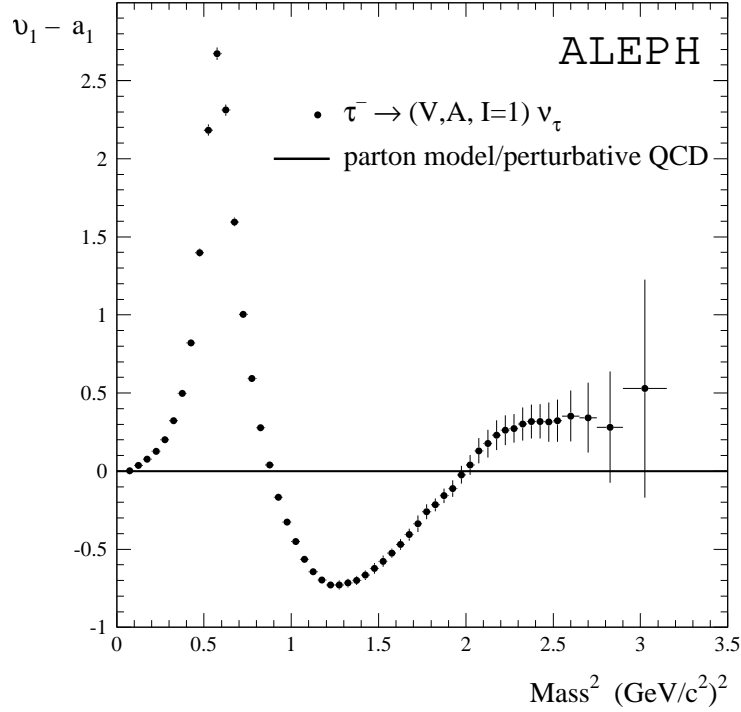}
\caption{Inclusive vector minus axial vector spectral function, $v_1-a_1$,
measured by the ALEPH collaboration \protect\cite{ALEPH2}.}
\label{f1}
\end{figure}
\begin{figure}[]
\includegraphics[width=11cm]{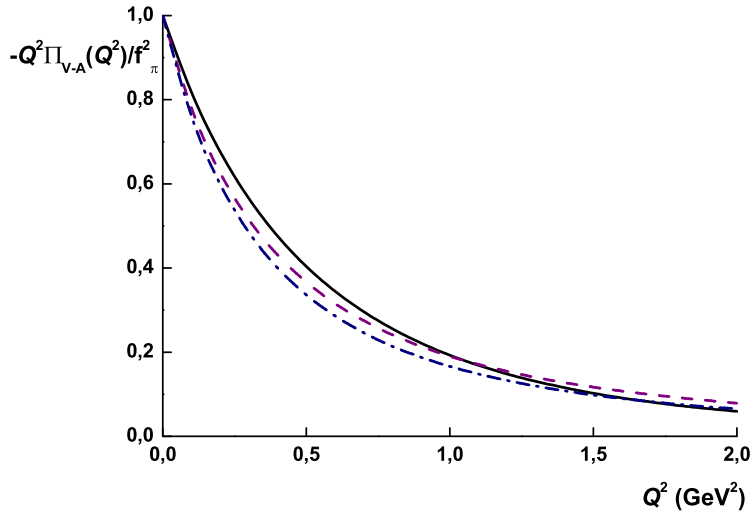}
\caption{Euclidean-momentum correlation function, $-Q^2
\Pi_{V-A}(Q^2)/f^2_\protect\protect\pi$, constructed in the present model
(solid line), in the model of Ref.~\protect\cite{deRafael:2002tj} (dashed
line), and reconstructed via Eq.~(\protect\ref{VmAreconstr}) from the ALEPH
experimental spectral function of Fig.~\protect\ref{f1} (dash-dotted line).}
\label{f2}
\end{figure}

\begin{figure}[]
\includegraphics[width=11cm]{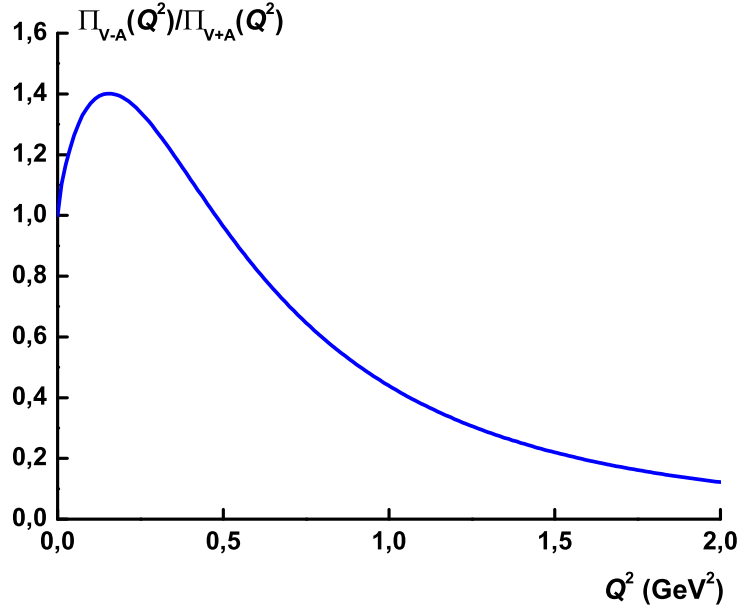}
\caption{The ratio of the nonperturbative parts of the $V-A$ to $V+A$
correlators in the nonsinglet channel.}
\label{f4}
\end{figure}

Next, we compare the model correlators with the ALEPH data, presented in
Fig. \ref{f1}. The ALEPH and OPAL data integrated up to the $\tau $ mass
satisfy all chiral sum rules within the experimental uncertainty, but the
central values differ significantly from the chiral model predictions.
Following Ref.~\cite{SHURYAK} we use $s_{0}=2.5 $ GeV$^{2}$ as an upper
integration limit, the value at which all chiral sum rules are satisfied
assuming that $v_{1}(s)-a_{1}(s)=0$ above $s_{0}$. Finally, a kinematic pole
at $q^{2}=0$ is added to the axial-vector spectral function. The resulting
unsubtracted dispersion relation between the measured spectral densities and
the correlation functions becomes
\begin{equation}
\Pi _{V}^{T}\left( Q^{2}\right) -\Pi _{A}^{T}\left( Q^{2}\right) =\frac{1}{%
4\pi ^{2}}\int_{0}^{s_{0}}ds\frac{v_{1}(s)-a_{1}(s)}{s+Q^{2}}-\frac{f_{\pi
}^{2}}{Q^{2}},  \label{VmAreconstr}
\end{equation}%
where $f_{\pi }^{2}$ is given by the WSR I,
\begin{equation}
f_{\pi }^{2}=\frac{1}{4\pi ^{2}}\int_{0}^{s_{0}}ds\left[ v_{1}(s)-a_{1}(s)%
\right] .
\end{equation}%
Having transformed the data into the Euclidean space, we may now proceed
with the comparison to the model, which obviously applies to the Euclidean
domain only. Admittedly, in the Euclidean presentation of the data the
detailed resonance structure corresponding to the $\rho $ and $a_{1}$ mesons
seen in the Minkowski region is smoothed out, hence the verification of
model results is not as stringent as would be directly in the Minkowski
space. In Fig. \ref{f2} we show the normalized curves corresponding to the
experimental data and the model prediction. We also show the prediction of
the model of Ref.~\cite{deRafael:2002tj} (minimal hadronic approximation,
MHA \cite{DeRaf98,Anri01})%
\begin{equation}
\left[ \Pi _{V-A}^{T}\left( Q^{2}\right) \right] _{MHA}=\frac{f_{\rho
}^{2}M_{\rho }^{2}}{Q^{2}+M_{\rho }^{2}}-\frac{f_{a}^{2}M_{a}^{2}}{%
Q^{2}+M_{a}^{2}}-\frac{f_{\pi }^{2}}{Q^{2}},  \label{PVmAMHA}
\end{equation}%
where the contributions of the $\rho $ and $a_{1}$ mesons are taken into
account with the model parameters $M_{\rho }=0.750$~GeV and $f_{a}=M_{\rho
}^{2}/M_{a}^{2}=0.5$. Other parameters are constrained by the Weinberg sum
rules. As demonstrated in Ref.~\cite{deRafael:2002tj}, the good agreement
between the data and the model predictions is far from trivial, since many
analytic approaches discussed in the literature meet definite difficulties
in the description of the ALEPH data in the region of moderately large $%
Q^{2} $. In Fig.~(\ref{f4}) we also present the ratio of the nonperturbative
parts of the $V-A$ (\ref{VmAmodel}) to $V+A$ correlators in the nonsinglet
channel.

To conclude this Section we wish to recall that quite similar calculations
of the vector and isovector axial-vector correlators within a nonlocal model
were done some time ago by Holdom and Lewis \cite{HoldLew95}. There are
certain differences in the form of the nonlocal interaction and, as a
consequence, the form of quark-current vertices is different. A more
principal difference is that the authors of \cite{HoldLew95} have chosen
a\textquotedblleft two phase\textquotedblright\ strategy, describing the
low-energy part of correlators by full nonperturbative vertices and
propagators, while the high energy parts were computed in the approximation
when one of the vertices is local. In this case the problem of matching of
two regimes occurs already at rather low energy scales. In the present
calculations one prolongs the applicability of the model up to moderately
large energies, which inter alias results in a good description of the ALEPH
data. On other side, we have to admit that one of the goals of both
approaches, namely the finding of a direct correspondence between the
effective model calculations and the OPE in QCD has not yet been reached
(see Sect. VII). To make correspondence closer it is necessary to supply the
model with a more detailed information on the soft quark-gluon interaction%
\footnote{%
Let us also refer to other important works \cite{YamawakiZakharov}, \cite%
{PerisRafael}, where the problem of connection between the effective
4-fermion models and QCD has been discussed.}.

\section{Current-current correlators (longitudinal parts)}

In this Section we demonstrate explicitly the transverse character of the $V$
and isovector (IV) $A$ correlators (Figs. (\ref{w9}) and (\ref{w10})). For
the longitudinal component of the $V$ correlator we get
\begin{equation}
K_{L}^{V}\left( Q^{2}\right) =\frac{4N_{c}}{Q^{2}}\int \frac{d^{4}k}{\left(
2\pi \right) ^{4}}\frac{M\left( k\right) }{D\left( k\right) }\left[ M\left(
k+Q\right) -M\left( k\right) \right] ,\qquad S_{L}^{V}\left( Q^{2}\right)
=-K_{L}^{V}\left( Q^{2}\right) ,
\end{equation}%
and therefore
\begin{equation}
-Q^{2}\Pi _{L}^{V}\left( Q^{2}\right) =0,  \label{PVL}
\end{equation}%
as it certainly should be by the requirement of the vector current
conservation.

Further, we consider the longitudinal projection of the $A$ correlator.
Then, we get contributions from the one-quark-loop diagram
\begin{equation}
K_{L,1}^{A}\left( Q^{2}\right) =-\frac{4N_{c}}{Q^{2}}\int \frac{d^{4}k}{%
\left( 2\pi \right) ^{4}}\frac{M\left( k\right) }{D\left( k\right) }\left[
M\left( k+Q\right) +M\left( k\right) \right] ,
\end{equation}%
the two-loop diagram in the isovector channel
\begin{equation}
K_{L,2}^{A}\left( Q^{2}\right) =\frac{8N_{c}}{Q^{2}}\left[ \int \frac{d^{4}k%
}{\left( 2\pi \right) ^{4}}\frac{M\left( k\right) }{D\left( k\right) }\sqrt{%
M\left( k+Q\right) M\left( k\right) }\right] ^{2}\left[ \int \frac{d^{4}k}{%
\left( 2\pi \right) ^{4}}\frac{M^{2}\left( k\right) }{D\left( k\right) }%
\right] ^{-1},
\end{equation}%
the one-loop contact diagram
\begin{equation}
S_{L,1}^{A}\left( Q^{2}\right) =\frac{4N_{c}}{Q^{2}}\int \frac{d^{4}k}{%
\left( 2\pi \right) ^{4}}\frac{M\left( k\right) }{D\left( k\right) }\left[
3M\left( k\right) +M\left( k+Q\right) -4\sqrt{M\left( k\right) M\left(
k+Q\right) }\right] ,
\end{equation}%
and, finally, from the two-loop contact diagram in the isovector channel%
\begin{equation}
S_{L,2}^{A}\left( Q^{2}\right) =-\frac{8N_{c}}{Q^{2}}\left[ \int \frac{d^{4}k%
}{\left( 2\pi \right) ^{4}}\frac{M\left( k\right) }{D\left( k\right) }\left[
M\left( k\right) -\sqrt{M\left( k+Q\right) M\left( k\right) }\right] \right]
^{2}\left[ \int \frac{d^{4}k}{\left( 2\pi \right) ^{4}}\frac{M^{2}\left(
k\right) }{D\left( k\right) }\right] ^{-1}.
\end{equation}%
The sum of all these contributions leads to the desired result%
\begin{equation}
-Q^{2}\Pi _{L}^{A,\mathrm{IV}}\left( Q^{2}\right) =0  \label{PAL}
\end{equation}%
consistent with the isovector axial current conservation in the strict
chiral limit.

\section{Singlet axial vector current correlator and the topological
susceptibility}

The cancellations in the longitudinal channels are consequences of the
current conservation and follow simply from the application of the
nonanomalous Ward-Takahashi identities. We have explicitly demonstrated this
in the previous Section in order to show the consistency of our
calculations. The issue becomes important when we consider the longitudinal
part of the isosinglet axial-vector current correlator which is not
conserved due to the $U_{A}\left( 1\right) $ axial Adler-Bell-Jackiw
anomaly. This channel is dominated not by the pion, but by the $\eta
^{\prime }$-meson intermediate state. Thus, in addition to the nonlocal
interaction present in Eq.~(\ref{Lint}) we also need to include the
interaction (\ref{G5}), where an exchange of the ``$\eta$'' singlet meson,
the $SU(2)$ analog of the $\eta ^{\prime }$ meson, occurs.

It is well known that due to the anomaly the singlet axial-vector current is
not conserved even in the chiral limit,
\begin{equation}
\partial _{\mu }J_{\mu }^{50}\left( x\right) =2N_{f}Q_{5}\left( x\right) ,
\end{equation}%
where $Q_{5}\left( x\right) $ is the topological charge density. In QCD it is defined as
$Q_{5}\left( x\right) =(\alpha_s/8\pi)G^a_{\mu\nu}(x)\tilde G^a_{\mu\nu}(x)$,
where $G^a_{\mu\nu}$ is the gluonic field strength, and $\tilde G^a_{\mu\nu}$ is its dual,
$\tilde G^a_{\mu\nu} = \varepsilon_{\mu\nu\lambda\sigma} G^a_{\lambda\sigma}$.
The correlator of the singlet axial-vector currents has the same Lorentz
structure as in Eq.~(\ref{PA}), but without flavor indices and with $%
T^{a}\equiv 1$. In the chiral limit its longitudinal part is related to the
topological susceptibility, the correlator of the topological charge
densities $Q_{5}\left( x\right) $,%
\begin{equation}
\chi \left( Q^{2}\right) =i\int d^{4}x~e^{iqx}\langle 0\left\vert T\left\{
Q_{5}(x)Q_{5}(0)\right\} \right\vert 0\rangle ,  \label{ChiQ2}
\end{equation}%
by the relation (see, \emph{e.g.}, \cite{IofSams00})%
\begin{equation}
\Pi _{L}^{A,0}\left( Q^{2}\right) =\frac{\left( 2N_{f}\right) ^{2}}{Q^{2}}%
\chi \left( Q^{2}\right) .  \label{PLchi}
\end{equation}%
At high $Q^{2}$ the OPE for $\chi\left( Q^{2}\right) $ predicts \cite{SVZeta}
\begin{equation}
\chi \left( Q^{2}\rightarrow \infty \right) =-\frac{\alpha _{s}}{16\pi }%
\left\langle \frac{\alpha _{s}}{\pi }\left( G_{\mu \nu }^{a}\right)
^{2}\right\rangle +\mathcal{O}(Q^{-2})+\mathcal{O}(e^{-Q\rho }),
\label{ChiQCDlarge}
\end{equation}%
where the perturbative contributions have been subtracted, and the
exponential corrections are due to nonlocal instanton interactions \cite{IofSams00}.

At low $Q^{2}$ the topological susceptibility $\chi \left( Q^{2}\right) $ can
be represented as a sum of contributions coming purely from QCD and from
hadronic resonances, \cite{IofSams00}. Crewther proved the theorem \cite{Crew}
that $\chi \left( 0\right) =0$ in any theory where at least one
massless quark exists (the dependence of $\chi \left( 0\right) $ on current
quark masses has been found in \cite{Venez79}\footnote{
Consistencies of the axial-vector current conservation and $U_{A}(1)$
problem is also discussed in \cite{Ohta80}.}). Also, the contributions of
nonsinglet hadron resonances are absent in the chiral limit. Thus, in the
low-$Q^{2}$ limit for massless current quarks one has
\begin{equation}
\left[ \chi \left( Q^{2}\rightarrow 0\right) \right] _{\chi QCD}=-Q^{2}\chi
^{\prime }(0)+\mathcal{O}(Q^{4}).  \label{ChiIof}
\end{equation}%
The estimates of $\chi ^{\prime }(0)$ existing in the literature are rather
controversial \footnote{%
The results obtained in Ref.~\cite{Fuku01} concerning $\chi (0)$ and $\chi
^{\prime }(0)$ contradict to the low-energy theorems.}:
\begin{equation}
\chi ^{\prime }(0)=\left( 48\pm 6~\mathrm{MeV}\right) ^{2}\qquad \lbrack
69],\qquad \chi ^{\prime }(0)=\left( 26\pm 4~\mathrm{MeV}\right) ^{2}\qquad
\lbrack 70].  \label{Chi1iof}
\end{equation}%
This makes further model estimates valuable.

Now we turn to the model calculations. The bare isosinglet axial-vector
current obtained from the interaction terms (\ref{NJLnl}) and (\ref{G5}) by
the rules described in Sect. IV becomes%
\begin{eqnarray}
\widetilde{\Gamma }_{\mu }^{50}(k,q,k^{\prime }=k+q) &=&\gamma _{\mu }\gamma
_{5}-\gamma _{5}(k+k^{\prime })_{\mu }\frac{\left( \sqrt{M(k^{\prime })}-%
\sqrt{M(k)}\right) ^{2}}{k^{\prime 2}-k^{2}}+  \label{G50bare} \\
&+&\gamma _{5}\frac{q_{\mu }}{q^{2}}2\sqrt{M(k^{\prime })M(k)}\left[ \frac{%
G^{\prime }}{M_{q}^{2}}J_{AP}(q^{2})-\frac{G^{\prime }}{G}\right] ,
\nonumber
\end{eqnarray}%
where $J_{AP}(q^{2})$is defined in Eq. (\ref{JAP}). In order to get the
full current we have to consider rescattering in the channel with the
quantum numbers of the singlet pseudoscalar meson, \textquotedblleft $\eta $%
\textquotedblright , which results in%
\begin{eqnarray}
\Gamma _{\mu }^{50}(k,q,k^{\prime }=k+q) &=&\gamma _{\mu }\gamma _{5}-\gamma
_{5}(k+k^{\prime })_{\mu }\frac{\left( \sqrt{M(k^{\prime })}-\sqrt{M(k)}%
\right) ^{2}}{k^{\prime 2}-k^{2}}  \label{G50} \\
&-&\gamma _{5}\frac{q_{\mu }}{q^{2}}2\sqrt{M(k^{\prime })M(k)}\frac{%
G^{\prime }}{G}\frac{1-GJ_{PP}(q^{2})}{1-G^{\prime }J_{PP}(q^{2})}.
\nonumber
\end{eqnarray}%
Because of the singlet axial anomaly this current does not contain the
massless pole anymore, since according to Eq. (\ref{PpiA0}) one has at zero
momentum:
\begin{equation}
\frac{1-GJ_{PP}(q^{2})}{-q^{2}}=G\frac{f_{\pi }^{2}}{M_{q}^{2}}.
\end{equation}%
Instead, it develops a pole at the \textquotedblleft $\eta $%
\textquotedblright\ meson mass. So, within the model considered the same mechanism
is responsible for the sponteneous breaking of chiral symmetry and violation of
the $U_A(1)$ symmetry.

The vertices satisfy the anomalous Ward-Takahashi identities:
\begin{equation}
q_{\mu }\widetilde{\Gamma }_{\mu }^{50}(k,q,k^{\prime }=k+q)=\widehat{q}%
\gamma _{5}-\gamma _{5}\left[ M(k^{\prime })+M(k)\right] +\gamma _{5}2\sqrt{%
M(k^{\prime })M(k)}\left( 1-\frac{G^{\prime }}{G}+G^{\prime }\frac{%
J_{AP}(q^{2})}{M_{q}^{2}}\right) ,
\end{equation}%
and%
\begin{equation}
q_{\mu }\Gamma _{\mu }^{50}(k,q,k^{\prime }=k+q)=\widehat{q}\gamma
_{5}-\gamma _{5}\left[ M(k^{\prime })+M(k)\right] +\gamma _{5}\frac{2\sqrt{%
M(k^{\prime })M(k)}}{1-G^{\prime }J_{PP}(q^{2})}\left( 1-\frac{G^{\prime }}{G%
}\right) ,  \label{AnWTI}
\end{equation}%
where the last term is due to the anomaly. Thus the QCD pseudoscalar
gluonium operator is interpolated by the pseudoscalar effective quark field
operator with coefficient expressed in terms of dynamical quark mass. In the
effective quark model the connection between quark and integrated gluon
degrees of freedom is fixed by the gap equation (\ref{SDEq})\cite{DP85}. By
considering the forward matrix element $(q=0)$ one deduces that the singlet
axial constant is not renormalized within our scheme: $G_{A}^{(0)}(0)=1$. In
order to get reduction of the singlet axial constant (\textquotedblleft spin
crisis\textquotedblright ) we need to consider the effects of polarization
of topologically neutral vacuum (see, \emph{e.g.}, \cite{Dorokhov:ym}).

It is instructive first to consider the longitudinal part of the correlator
of the local vertex, $\gamma _{\mu }\gamma _{5}$, and the nonlocal vertex of
Eq.~(\ref{G50}), which is the construction of Pagels and Stokar \cite%
{pagsto79}. In this model the decay constant is defined by%
\begin{equation}
f_{\pi ,PS}^{2}=\frac{N_{c}}{8\pi ^{2}}\int\limits_{0}^{\infty }du\ u\frac{%
2M(u)^{2}-uM(u)M^{\prime }(u)}{D^{2}\left( u\right) }.  \label{Fpi2_PagSt}
\end{equation}%
Then, through the use of Eq.~(\ref{PLchi}) we get for the topological
susceptibility the result
\begin{equation}
\left( 2N_{f}\right) ^{2}\chi _{PS}\left( Q^{2}\right) =-\frac{G-G^{\prime }%
}{G\left[ 1-G^{\prime }J_{PP}(Q^{2})\right] }\frac{N_{c}N_{f}}{4\pi ^{2}}%
\int d^{4}k\frac{\sqrt{M_{+}M_{-}}}{D_{+}D_{-}}\left[ M_{+}\left(
k_{-}q\right) -M_{-}\left( k_{+}q\right) \right] ,  \label{ChiPagStok}
\end{equation}%
with the coefficients of the low-$Q^{2}$ expansion given by%
\begin{equation}
\chi _{PS}\left( 0\right) =0,\qquad \chi _{PS}^{\prime }\left( 0\right) =%
\frac{f_{\pi ,PS}^{2}}{2N_{f}} .  \label{Chi1pagStok}
\end{equation}%
Hence, the result is consistent with the Crewther theorem and it provides
the estimate of $\chi _{PS}^{\prime }\left( 0\right) \approx \left( 39~%
\mathrm{MeV}\right) ^{2}$ obtained for $N_f=3$. The second relation may be
rewritten in the form resembling the generalized Goldberger-Treiman
relation, as advocated by Veneziano and Shore \cite{NShVenez}. Indeed, by
using the standard Goldberger-Treiman relation, (\ref{Gpi}), valid in a
given model, one finds
\begin{equation}
M_{q}=g_{\pi }^{PS}\sqrt{2N_{f}\chi _{PS}^{\prime }\left( 0\right) },
\label{GTRg}
\end{equation}%
which is just the quark-level relation from Ref.~\cite{NShVenez}. At large $%
Q^{2}$ the quantity $\chi _{PS}\left( Q^{2}\right) $ decreases according to
the power of the dynamical quark form factor.

Now let us turn to the full model calculation. Proceeding in a similar
manner as in the previous Sections we get the topological susceptibility in
the form
\begin{eqnarray}
-\left( 2N_{f}\right) ^{2}\chi \left( Q^{2}\right) &=&2N_f\left( 1-\frac{%
G^{\prime }}{G}\right) \left\{ Q^{2}J_{\pi A}\left( Q^{2}\right) \left[ 1-%
\frac{G^{\prime }J_{AP}(Q^{2})}{M_{q}^{2}}+\frac{1}{1-G^{\prime
}J_{PP}(Q^{2})}\right] \right.  \label{ChiMod} \\
&+&M_{q}^{2}J_{PP}\left( Q^{2}\right) \left( 1-\frac{G^{\prime }}{M_{q}^{2}}%
J_{AP}(Q^{2})\right) \left[ \frac{GJ_{AP}(Q^{2})}{M_{q}^{2}}-\frac{%
G-G^{\prime }}{G\left[ 1-G^{\prime }J_{PP}(Q^{2})\right] }\right]  \nonumber
\\
&+&\left. \frac{G}{M_{q}^{2}}\left[ 4N_{c}N_{f}\int \frac{d^{4}k}{\left(
2\pi \right) ^{4}}\frac{M\left( k\right) }{D\left( k\right) }\left[ M\left(
k\right) -\sqrt{M\left( k+Q\right) M\left( k\right) }\right] \right]
^{2}\right\} ,  \nonumber
\end{eqnarray}%
where we have used the relation between the singlet current correlator and
the topological susceptibility of Eq. (\ref{PLchi}). At large $Q^{2}$ one
obtains the power-like behavior consistent with the OPE prediction (\ref%
{ChiQCDlarge}), namely
\begin{equation}
-\left( 2N_{f}\right) ^{2}\chi \left( Q^{2}\rightarrow \infty \right) =\frac{%
2N_fM_{q}^{2}}{G}\left( 1-\frac{G^{\prime }}{G}\right) .  \label{Chi1As}
\end{equation}%
At zero momentum the topological susceptibility is zero
\begin{equation}
\chi \left( 0\right) =0,
\end{equation}%
in accordance with the Crewther theorem. For the first moment of the
topological susceptibility we obtain
\begin{equation}
\chi ^{\prime }\left( 0\right) =\frac{1}{2N_{f}}\left\{ f_{\pi }^{2}\left( 2-%
\frac{G^{\prime }}{G}\right) +\left( 1-\frac{G^{\prime }}{G}\right)
^{2}J_{AP}^{\prime }(0)\right\} ,  \label{Chi1mod}
\end{equation}%
where $f_{\pi }^{2}$ and $J_{AP}^{\prime }(0)$ are defined in Eqs.~(\ref%
{Fpi2_M}) and (\ref{J1AP0}), respectively.

In order to get numerical results we need to specify further the details of
the model. We consider two possibilities. One involves the interaction with
the exact symmetry as provided by the 't Hooft determinant, $G^{\prime }=-G$%
. For the second possibility the constants $G$ and $G^{\prime }$ are
considered as independent of each other, and their values are fixed with the
help of the meson spectrum. In this more realistic scenario one has
approximately the relation $G^{\prime }=0.1~G$ (for typical sets of
parameters, \emph{c.f.} Ref.~\cite{Birse95}). Then we get the following
estimates for the first moment of the topological susceptibility:
\begin{eqnarray}
\chi ^{\prime }(0) &=&\left( 55~\mathrm{MeV}\right) ^{2}\qquad (G^{\prime
}=-G), \\
\chi ^{\prime }(0) &=&\left( 50~\mathrm{MeV}\right) ^{2}\qquad (G^{\prime
}=0.1~G).  \label{Ch1MN}
\end{eqnarray}%
For the above estimates we have taken $N_{f}=3$. Since the flavor number
enters only through the factor of $2N_{f}$ present in the definition (\ref%
{PLchi}), in this sense it is external to the model and its inclusion is
very simple. We can see that the model gives the values of $\chi ^{\prime
}(0)$ which are close to the estimate of Ref.~\cite{IoOg98}. In Fig. 3 we
present the model predictions for the topological susceptibility at low and
moderate values of $Q^{2}$ for the cases of the full (\ref{ChiMod}) and
Pagels-Stokar (\ref{ChiPagStok}) model calculations.

We should note that the predictions of our model have a limited character
because we have used the $SU_{f}\left( 2\right) $ model in the chiral limit
and have not considered mixing effects. However, our final result is
formulated in terms of a physical observable, $f_{\pi }$, and thus we
believe that the presented predictions may be not far from more realistic
model calculations. In the region of small and intermediate momenta our
results are quantitatively close to the predictions of the QCD sum rules
with the instanton effects included \cite{IofSams00}. In our opinion, both
the instanton-based calculations (our model with $\left( G^{\prime
}=-G\right) $ and the interpolation of the model \cite{IofSams00})
overestimate the instanton contributions in the region $Q^{2}\sim 0.5-2$ GeV$%
^{2}$. It would be interesting to verify the predictions given in Fig. \ref%
{f3} by the modern lattice simulations.

\begin{figure}[]
\includegraphics[width=10cm]{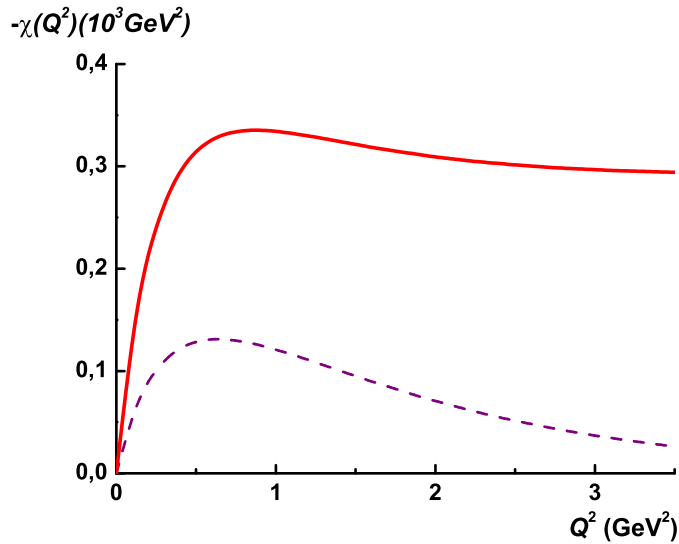}
\caption{Topological susceptibility versus $Q^2$ predicted by the model with
$G^{\prime}=0.1~G$, Eq. (\protect\ref{ChiMod}), (solid line), and by the
Pagels-Stokar construction, Eq. (\protect\ref{ChiPagStok}) ,(dashed line).}
\label{f3}
\end{figure}

\section{Conclusions}

In this work we have analyzed the nonperturbative parts of the
Euclidean-momentum correlation functions of the vector and axial-vector
currents within an effective nonlocal chiral quark model. To this end, we
have derived the conserved vector and isotriplet axial-vector currents and
demonstrated explicitly the absence of longitudinal parts in the $V$ and
nonsinglet $A$ correlators, which is consequence of the gauge invariance of
the present approach. On the other hand, the singlet $A$ correlator gains an
anomalous contribution. From the properties of the $V-A$ correlator we have
shown the fulfillment of the low-energy relations. The values of the $\pi
^{\pm }-\pi ^{0}$ electromagnetic mass difference and the electric pion
polarizability are estimated and found to be remarkably close to the
experimental values. In the high-energy region the relation to OPE has been
discussed. In particular, the estimate of the $1/Q^{2}$ coefficient is in
agreement with the recent lattice findings and the modified OPE
phenomenology. We stress that the momentum dependence of the dynamical quark
mass is crucial for the fulfillment of the second Weinberg sum rule. The
combination $V-A$ receives no contribution from perturbative effects and
provides a clean probe for chiral symmetry breaking and a test ground for
model verification. We have found that our model describes well the
transformed data of the ALEPH collaboration on the hadronic $\tau $ decay.
The combination $V+A$, on the other hand, is dominated by perturbative
contributions which are subtracted from our analysis. By considering the
correlator of the singlet axial-vector currents the topological
susceptibility has been found as a function of the momentum, and its first
moment is predicted. In addition, the fulfillment of the Crewther theorem
has been demonstrated.

\begin{acknowledgments}
AED is grateful to S. B. Gerasimov, S.V. Mikhailov, N. I. Kochelev, M. K.
Volkov, A. E. Radzhabov, L. Tomio, and H. Forkel for useful discussions of
the subject of the present work. WB thanks M. Polyakov and E. Ruiz Arriola
for discussions on certain topics at an early stage of this research. AED
thanks for partial support from RFBR (Grants nos. 01-02-16431, 02-02-16194,
03-02-17291), INTAS (Grant no. 00-00-366), and the \textquotedblleft
Fundacao de Amparo ~ Pesquisa do Estado de Sao Paulo
(FAPESP)\textquotedblright . We are grateful to the Bogoliubov-Infeld
program for support.
\end{acknowledgments}


\end{document}